\documentclass[12pt,a4paper,aps,prd,preprint,superscriptaddress,nofootinbib]{revtex4-1}
\usepackage[utf8]{inputenc}
\usepackage{graphicx}
\usepackage{amssymb}
\usepackage{textcomp}
\usepackage{amsmath}
\usepackage{tabularx}
\usepackage{bm}
\usepackage{times}
\usepackage{color}
\usepackage{slashed}
\usepackage{multirow}
\usepackage{verbatim}
\usepackage{cancel}
\usepackage{subfigure}
\usepackage[normalem]{ulem}

\usepackage[colorlinks=true, pdfstartview=FitV, linkcolor=blue, citecolor=blue, urlcolor=blue]{hyperref}
\allowdisplaybreaks[4]

\def\lsim{\mathrel{\raise.3ex\hbox{$<$\kern-.75em\lower1ex\hbox{$\sim$}}}}
\def\gsim{\mathrel{\raise.3ex\hbox{$>$\kern-.75em\lower1ex\hbox{$\sim$}}}}

\newcommand{\GeV}{{\rm GeV}}

\newcommand{\bew}{\begin{widetext}}
\newcommand{\enw}{\end{widetext}}
\newcommand{\bee}{\begin{equation}}
\newcommand{\ene}{\end{equation}}
\newcommand{\bea}{\begin{eqnarray}}
\newcommand{\ena}{\end{eqnarray}}
\newcommand{\bes}{\begin{subequations}}
\newcommand{\ens}{\end{subequations}}

\def\calc{\mathcal{C}}

\def\calm{\mathcal{M}}

\definecolor{deepPurple}{RGB}{150,100,200}
\definecolor{orange}{rgb}{1,0.5,0}

\begin{document}

\title{Probing the chromo-electric dipole moment of top quark near the toponium production threshold at LHC}

\author{Kai Ma}
\email{kai@xauat.edu.cn}
\affiliation{Faculty of Science, Xi'an University of Architecture and Technology, Xi'an, 710055, China}

\author{Tong Li}
\email{litong@nankai.edu.cn}
\affiliation{
School of Physics, Nankai University, Tianjin 300071, China
}

\begin{abstract}
The recent observation of an excess near the kinematic $t\bar{t}$ production threshold by the CMS and ATLAS collaborations suggests the existence of a color-singlet and $S$-wave quasi-bound state of top quark. Due to the non-relativistic nature of top and anti-top quarks in this threshold region, toponium production provides an ideal system to probe the anomalous dipole moments of the top quark. In this work, we investigate the impact of chromo-electric dipole moment (CEDM) in $t\bar{t}$-gluon vertex on toponium production at the LHC. We implement the enhanced near-threshold production of the spin-0 color-singlet toponium state $\eta_t$ via gluon fusion. The resonant toponium contribution is then combined
with the continuum $t\bar{t}$ production amplitude, including the interference effects induced by the top CEDM. Furthermore, we construct a set of asymmetries based on the angular distributions of charged leptons in the leptonic decay channel of $t\bar{t}$. Finally, we derive the expected exclusion limit on the top quark CEDM parameter $d_t^g$ using these lepton angular asymmetries.
\end{abstract}

\maketitle
\tableofcontents

\section{Introduction}
\label{sec:Intro}

Quarkonium as the bound state of heavy quark and anti-quark is one of the most important study aspects of Quantum Chromodynamics (QCD). Over a long time, the top quark as the heaviest quark in the Standard Model (SM) has been widely accepted to be too short-lived ($\Gamma_t\approx 1.42$ GeV, $\tau_t\approx 10^{-25}$ s) to form a bound state and thus there are no any toponia~\cite{Griffiths:1987tj}. Nevertheless, the production of a pair of top and anti-top quarks near threshold was early suggested in $e^+ e^-$ collision~\cite{Fadin:1987wz}. Although there were some theoretical studies of toponium in non-relativistic QCD (NRQCD) framework later~\cite{Fadin:1990wx,Strassler:1990nw,Fadin:1991zw,Sumino:1992ai,Jezabek:1992np,Fabiano:1993vx,Fadin:1994pj,Hagiwara:2008df,Kiyo:2008bv,Sumino:2010bv,Ju:2020otc,Fuks:2021xje,Garzelli:2024uhe,Llanes-Estrada:2024phk,Shao:2025dzw}, the observation of toponium at hadron colliders such as the Large Hadron Collider (LHC) was still in debate.

In recent years, the LHC as a top factory did show some hints of disagreement between the measured observables and predictions near the $t\bar{t}$ kinematic production threshold~\cite{CMS:2018adi,ATLAS:2023fsd,CMS:2024pts,ATLAS:2025doc}. More recently, both CMS~\cite{CMS:2025kzt} and ATLAS~\cite{ATLAS:2026nrx} reported the enhancement of resonant $t\bar{t}$ production near the threshold. A statistical significance exceeding $5\sigma$ deviation was observed. This excess is consistent with color-singlet and $S$-wave $t\bar{t}$ quasi-bound state. The CMS analysis, based on $138~{\rm fb}^{-1}$ of $pp$ data at $\sqrt{s}=13$ TeV, reported a significant excess near the kinematic $t\bar t$ threshold consistent with a color-singlet pseudoscalar quasi-bound state and measured an excess cross section of $8.8^{+1.2}_{-1.4}$ pb.
The ATLAS analysis, using $140~{\rm fb}^{-1}$ at the same collision energy, observed an excess over the baseline perturbative QCD prediction with significance above $8\sigma$, corresponding to a cross section of $9.3^{+1.4}_{-1.3}$ pb.
These results strongly motivate a dedicated theoretical treatment of toponium formation as part of the SM prediction for $t\bar t$ production near threshold. There then appeared dedicated studies of toponium's formation~\cite{Fuks:2024yjj,Aguilar-Saavedra:2024mnm,Francener:2025tor,Fuks:2025sxu,Xiong:2025iwg,Fuks:2025wtq,Fuks:2025toq,Li:2026nmr,Garzelli:2026ctb,Garzelli:2026ctb} and relevant phenomena~\cite{Li:2022iav,Fu:2025yft,Bai:2025buy,Luo:2025psq,Najjar:2025bby,Flacke:2025dwk,Antozzi:2026vdi,Zhang:2026iuw,Aguilar-Saavedra:2026grl,Yang:2026enl,Jia:2026dui,Rodriguez:2026pfy}.
The Bohr radius of the toponium state can be given by the Coulomb-like QCD potential as $a_0\approx (20~{\rm GeV})^{-1}$~\cite{Bai:2025buy,Fuks:2025sxu} which is much smaller than the boundary radius of asymptotic freedom and confinement. Its decay width becomes $\Gamma_t+\Gamma_{\bar{t}}=2\Gamma_t\approx 2.84$ GeV.

On the other side, the study of new physics beyond the SM in top quark sector spans several decades.
The origin of CP symmetry violation is of great interest in particle physics. The search for CP violation would reveal the new physics relevant to both the origin of matter mass and the matter-antimatter asymmetry in the universe. Due to the short lifetime of top quark, it is generally assumed that the CP violation phenomena in the top sector does not receive impact from the non-perturbative bound state physics. The seek for CP violation in top quark as an elementary fermion has no complications in such as $K$ meson or $B$ meson system. It is thus well motivated to study CP violation in the top quark system (see Ref.~\cite{Atwood:2000tu} for a recent review and references therein).
There existed plenty of studies about the CP properties of the observed 125 GeV Higgs boson $h$ in the production of $t\bar{t}h$ governed by the large top Yukawa coupling.
The CP violating anomalous top quark couplings can also arise due to a dipole moment in the $t\bar{t}$-gluon or $t\bar{t}$-$Z/\gamma$ vertex of the tensor current form $\bar{t}\sigma_{\mu\nu}t$ or $\bar{t}\sigma_{\mu\nu}\gamma_5 t$. In the extensions of the SM, such as multi-Higgs doublet models~\cite{Weinberg:1976hu,Albright:1979yc,Shizuya:1980jx,Bernreuther:1992dz,Chang:1993fu} or supersymmetric models~\cite{Christova:1992ee,Bernreuther:1993xp,Grzadkowski:1993kb,Bartl:1995gj,Bartl:1997iq}, the top dipole moment can be 10 orders of magnitude larger than the SM prediction. Potential asymmetric observables were proposed to probe CP violating anomalous top quark couplings in gauge sector at high-energy hadron colliders or lepton colliders~\cite{Donoghue:1987ax,Ma:1992nyl,Bernreuther:1992be,Brandenburg:1992be,Atwood:1992vj,Bernreuther:1993hq,Choi:1997ie,Zhou:1998wz,Antipin:2008zx,Gupta:2009wu,Bernreuther:2017cyi,Isildak:2023dnf}. The relevant search is most noticeable in the high invariant mass ($m_{t\bar{t}}$) regime far above the kinematic threshold for the production of elementary top and anti-top quarks~\cite{CMS:2019kzp}. There was no dedicated study of CP violating anomalous top quark couplings near the kinematic threshold.

Motivated by the observation of toponium, it is natural to examine the implication of toponium formation for CP violation near the kinematic threshold of $t\bar{t}$ production. The threshold region provides the lowest momentum scale accessible in direct on-shell $t\bar t$ production, due to the non-relativistic nature of top and anti-top quarks. The above measurements for elementary top and anti-top quarks would be likely not applicable in this region. It is worth examining the effect of anomalous top couplings near threshold in details. There was the probe of CP violation in the Higgs boson and toponia production at the $e^+ e^-$ collider~\cite{Hagiwara:2016rdv}.
In this work, we investigate the impact of chromo-electric dipole moment (CEDM) in $t\bar{t}$-gluon vertex on toponium production at the LHC.
The CP violating interaction between top quark and gluon from CEDM is of the form
\begin{eqnarray}
\overline{t}\sigma_{\mu\nu}\gamma_5t G^{\mu\nu}\;,
\end{eqnarray}
where $G^{\mu\nu}$ denotes the gluon field strength. The inclusion of CEDM modifies the $gt\bar{t}$ vertex and introduces a new $ggt\bar{t}$ vertex.
The near-threshold toponium production at the LHC is expected to be dominantly generated by gluon fusion channel.
The subdominant channels arise from quark-antiquark annihilation via the exchange of gluon, photon and $Z$ boson.
In general, the near-threshold $t\bar t$ system can be decomposed into states of different color representations.
However, only the production of color-singlet state receives an attractive QCD potential and can lead to a resonantly enhanced quasi-bound-state contribution~\cite{Fadin:1990wx,Strassler:1990nw,Fadin:1991zw,Sumino:1992ai,Jezabek:1992np,Fabiano:1993vx,Fadin:1994pj,Hagiwara:2008df,Kiyo:2008bv,Sumino:2010bv}. The color-octet state is not expected to induce such an enhancement.
Moreover, the leading near-threshold contribution comes from the $S$-wave states, whereas higher partial waves are suppressed by powers of the relative velocity of the top quarks.
In this work, we therefore consider only the non-perturbative enhancement in the gluon fusion production of spin-0 color-singlet $S$-wave state. We will also show that the contribution of spin-1 color-singlet $S$-wave state from quark-antiquark annihilation mediated by $\gamma/Z$ is numerically negligible. For the color-octet $S$-wave states from gluon fusion and $q\bar q$ annihilation, we will not implement the NRQCD Green's function due to their essentially small contribution in the near-threshold regime. The $P$-wave states will not be considered in our analysis because of the suppressed Green's function.

The non-relativistic bound-state dynamics are encoded in the toponium line shape by an overall production factor. The spin correlations of the final top and anti-top quarks are kept in their subsequent decays.
This characteristic allows us to combine the resonant toponium contribution with the continuum $t\bar t$ production amplitude and to consistently include the interference effects induced by the top CEDM.
To probe the CP-violating interaction, we further construct a set of asymmetries from the angular distributions of the charged leptons in the leptonic decay channel of $t\bar{t}$.
Following the spin-correlation method in Refs.~\cite{Ma:2017vve,Ma:2018ott}, these asymmetries are defined as
\begin{eqnarray}
A_{\cal \xi}=
{N({\xi}>0)-N({\xi}<0)\over
N({\xi}>0)+N({\xi}<0)}\;,
\label{eq:asym}
\end{eqnarray}
where $\xi$ denotes suitable observable by combining the polar and/or azimuthal angles of $\ell^+$ and $\ell^-$ in a chosen reference frame.
Conventional analyses often define the lepton angular observables in the rest frames of the reconstructed top and anti-top quarks, or in the $t\bar t$ rest frame.
However, because of finite detector resolution and the large missing energy carried by neutrinos in leptonic $t\bar t$ events, such observables can be significantly affected by reconstruction uncertainties.
We examine the spin correlations in a single reference frame using a helicity-projected basis. Fifteen independent observables $\xi_i~(i=1-15)$ are defined in this basis.
The observables adopted in our approach can be calculated in any desired reference frame, including the frames that do not require the full reconstruction of the top and anti-top momenta. Since charged leptons are excellent analyzers of the top quark spin, these asymmetries provide direct access to the polarization and spin-correlation structures of the produced $t\bar t$ system.
In particular, the CP-odd asymmetries vanish in the absence of CP violation up to tiny SM effects. However, they receive linear contributions from the interference between the amplitudes with and without CEDM.
They therefore serve as sensitive probes of anomalous CP violation in toponium production near the kinematic threshold.

This paper is organized as follows.
In Sec.~\ref{sec:CP}, we review the CP violating interaction of top quark and gluon.
We present the detailed calculations of the toponium production at the LHC in Sec.~\ref{sec:toponium}. The asymmetries in toponium production with CEDM are discussed in Sec.~\ref{sec:Res}.
We also show the prospective sensitivity of toponium production to top CEDM at the LHC. We draw our conclusions in Sec.~\ref{sec:Con}.

\section{Chromo-electric dipole moment of top quark}
\label{sec:CP}

The dipole interactions of top quark have been extensively studied both theoretically and
experimentally.
In the Warsaw-basis notation of SM effective field theory (SMEFT)~\cite{Grzadkowski:2010es}, the
dimension-six dipole operator for top quark and gluon is
\begin{eqnarray}
{\cal O}_{uG}^{33} &=&
(\bar Q_3\sigma^{\mu\nu}T^A t_R)\tilde\varphi\,G^A_{\mu\nu}\;,
\end{eqnarray}
where $Q_3$ is the third-generation left-handed quark doublet, $t_R$ is the
right-handed top quark, $\tilde\varphi=i\tau^2\varphi^\ast$ with $\varphi$ being the SM Higgs doublet, $G^A_{\mu\nu}$ is the field-strength tensor
of $SU(3)_c$, and $T^A$ denotes the $SU(3)_c$ generator. After electroweak symmetry
breaking with $\varphi$ receiving the SM vacuum expectation value (vev) $\langle \varphi\rangle=v_0/\sqrt{2}$, the operator ${\cal O}_{uG}^{33}$ generates the anomalous dipole interactions between top quark and gluon as follows
\begin{eqnarray}
{\cal L}_{t\bar t g}^{\rm dipole}
=
{v_0\over \sqrt{2}\Lambda^2}
\bar t\,\sigma^{\mu\nu}T^A
\left[
{\rm Re}\,C_{uG}^{33}
+i\,{\rm Im}\,C_{uG}^{33}\gamma_5
\right]t\,G^A_{\mu\nu}\;,
\end{eqnarray}
where $C_{uG}^{33}$ is the dimensionless Wilson coefficient of ${\cal O}_{uG}^{33}$ and $\Lambda$ denotes the ultraviolet (UV) cutoff scale~\footnote{In the notation of Ref.~\cite{Hayreter:2015ryk}, there is an additional strong coupling constant $g_s$ in the definition of SMEFT dipole interaction.}.
The general top-gluon dipole interaction can be parameterized as~\cite{Atwood:1994vm,Haberl:1995ek,Cheung:1995nt,Choi:1997ie}~\footnote{\label{fnt:sign}The sign notations differ across some of these references.}
\begin{eqnarray}
{\cal L}_{t\bar t g}^{\rm dipole}
=
-
{g_s\over 2m_t}
\bar t\,T^A\sigma^{\mu\nu}
\left(
d_V^g+i d_A^g\gamma_5
\right)t\,G^A_{\mu\nu}\;,
\end{eqnarray}
where $d_V^g$ and $d_A^g$ denote the CP-even chromo-magnetic dipole moment (CMDM) and CP-odd chromo-electric dipole
moment (CEDM), respectively. The other usual parameterization is~\cite{Brandenburg:1992be,Atwood:1992vj,Haberl:1995ek,Hayreter:2015ryk}~\footnotemark[\getrefnumber{fnt:sign}]
\begin{eqnarray}
{\cal L}_{t\bar t g}^{\rm dipole}
=
-
{g_s\over 2}
\bar t\,T^A\sigma^{\mu\nu}
\left(
a_t^g+i d_t^g\gamma_5
\right)t\,G^A_{\mu\nu}\;,
\end{eqnarray}
where $a_t^g=d_V^g/m_t$ and $d_t^g=d_A^g/m_t$. We will adopt the last parameterization in our analysis.

The SMEFT framework provides a gauge-invariant
description of these dipole interactions and allows their effects to be consistently included in
top-pair production, single-top production and top decay~\cite{Aguilar-Saavedra:2008nuh,Zhang:2010dr}.
The CP-even CMDM modifies the total rate and differential
distributions of $t\bar t$ production, while the CP-odd CEDM gives
rise to CP-odd momentum and spin correlations~\cite{Bernreuther:1992dz,Atwood:2000tu,Bernreuther:2013aga,Hayreter:2015ryk}.
In the SM, the CP violation in top quark sector is through the exchange of $W$ boson at one-loop level and is expected
to be extremely small. An observable top CEDM would indicate a new source of CP
violation. Indirect constraints can arise from low-energy EDM
measurements through renormalization group running and matching onto the EDMs or CEDMs of light-quark, and the CP-odd three-gluon Weinberg operator~\cite{Kamenik:2011dk,Brod:2013cka,Dekens:2014jka}.
These indirect bounds can be numerically strong, but they highly depend on the assumed UV
completion, flavor structure, operator mixing, and possible cancellations among different
CP-violating operators. Direct collider measurements are therefore indispensable.
The forward-backward asymmetry of top quark, along with the anomalous CMDM and CEDM, has been measured using $t\bar t$ events at $\sqrt{s}=13$ TeV LHC~\cite{CMS:2019kzp}. The event data are selected for large reconstructed mass of top quark pair with $m_{t\bar{t}}>350$ GeV. Based on
Lorentz-scalar and momentum-correlation asymmetries, the dedicated searches
for CP violation in leptonic and semi-leptonic $t\bar t$ events are consistent with the
SM predictions and yield no evidence of a nonzero top CEDM~\cite{CMS:2019kzp,CMS:2022voq,CMS:2022quh}.
The current LHC upper limit on the CEDM of top quark at 95\% confidence level (CL) is~\cite{CMS:2022voq}
\begin{eqnarray}
\label{eq:bul:CMS}
\left| d_t^g \right|_{\rm Exp.}<0.015 /m_t \approx 8.696\times 10^{-5}~\GeV^{-1}
\,.
\end{eqnarray}

Nevertheless, the near-threshold $t\bar t$ production provides a unique probe of dipole interactions because of the minimum hard momentum scale, of order $m_t$, allowed by the kinematics of direct on-shell $t\bar{t}$ production.
Near the production threshold,
the top and anti-top quarks are non-relativistic and their relative velocity $v$ is small. This implies that their energy-momentum becomes $|\vec p\,|\sim m_t v$ and $E\sim m_t v^2$ with $v\ll 1$.
These low scales characterize the long-distance dynamics of the $t\bar t$ system and can be systematically separated from the short-distance production process. The threshold region therefore provides a particularly well-defined setting for probing the dipole interactions of top quark at the
lowest hard scale accessible at a collider.
Furthermore, for CEDM, the dominant
contributions come from interference between the SM amplitude and the CEDM-induced
amplitude. They can generate CP-odd spin correlations which are linear in ${\rm Im}\,C_{uG}^{33}$, or equivalently in $d_t^g$.
In this work, we therefore
focus on the chromo-electric dipole interaction of the top quark and investigate its
impact on CP-violating asymmetry observables in toponium production near the
$t\bar t$ kinematic threshold. We will use the following effective CEDM Lagrangian of top quark and gluon
\begin{eqnarray}
\mathcal{L}^{\rm CEDM}_{t\bar{t}g}=-ig_s {d_t^g\over 2} \overline{t} T^A\sigma^{\mu\nu}\gamma_5 t G_{\mu\nu}^A \;.
\end{eqnarray}
The Feynman rules of corresponding vertex are given by~\cite{Antipin:2008zx}
\begin{eqnarray}
&&gt\overline{t}: -ig_s T^A\Big(\gamma^\mu + d_t^g \sigma^{\mu\nu} q_\nu \gamma_5\Big)G^A_\mu\;,\\
&&ggt\overline{t}: i g_s^2[T^B,T^C] d_t^g \sigma^{\mu\nu} \gamma_5 G^B_\mu G^C_\nu\;,
\end{eqnarray}
where $q$ denotes the incoming momentum of gluon.
We build the Universal FeynRules Output (UFO)~\cite{Degrande:2011ua} files of CEDM and pass them together with SM model files to MadGraph5\_aMC@NLO~\cite{Alwall:2014hca} to generate the signal events at parton-level. The mass and width of the top quark are set to be $173\,$GeV and $1.49\,$GeV, respectively~\cite{Fuks:2025wtq}.

It is important to note that the existing collider constraints on the top CEDM are generally extracted from observables characterized by different kinematic and renormalization scales. In particular, the bound in Eq.~\eqref{eq:bul:CMS} should not be directly interpreted as a constraint on the CEDM at the $t\bar t$ threshold without properly accounting for the corresponding scale dependence. The threshold region provides the lowest hard scale accessible in direct on-shell $t\bar t$ production, of order $m_t$. We therefore take $d_t^g = 3\times10^{-4}~\GeV^{-1}$ as a benchmark value in the following discussion to illustrate the effects induced by the top CEDM, rather than as a value inferred directly from the bound in Eq.~\eqref{eq:bul:CMS}.

\section{Toponium production at the LHC}
\label{sec:toponium}

At a hadron collider, the toponium production should not be regarded as a process completely separated from ordinary top pair production.
Rather, it corresponds to the threshold enhancement of the same $t\bar t$ production amplitudes when the relative velocity of the top and anti-top quarks becomes small.
For a partonic collision with invariant mass $m_{t\bar t}$, it is convenient to introduce the non-relativistic energy variable
\begin{eqnarray}
E=m_{t\bar t}-2m_t\;,
\qquad
v\simeq \sqrt{E/m_t}\;,
\end{eqnarray}
where the relative velocity $v\ll 1$ in the threshold region.
In this regime, the ordinary fixed-order expansion in $\alpha_s$ becomes insufficient. This is because the exchange of potential gluons between the slowly moving top and anti-top quarks generates Coulomb-enhanced terms of the form $(\alpha_s/v)^n$.
These contributions have to be resummed.
At the same time, the large top quark decay width $\Gamma_t$ provides a physical infrared regulator and smears the would-be discrete bound state spectrum into a broad enhancement near the $t\bar t$ threshold.
Therefore, toponium at the LHC is more appropriately described as a short-lived $t\bar t$ quasi-bound state in the near-threshold line shape, rather than as a narrow quarkonium resonance in the conventional sense~\cite{Fadin:1987wz,Fadin:1990wx,Strassler:1990nw,Kiyo:2008bv,Sumino:2010bv,Fuks:2021xje}.

The hadronic invariant mass distribution of $t\bar t$ production can be written schematically as
\begin{eqnarray}
{d\sigma(pp\to t\bar t X)\over dm_{t\bar t}}
=
{2m_{t\bar t}\over s}
\sum_{i,j}
{\cal L}_{ij}\left({m_{t\bar t}^2\over s},\mu_F\right)
\hat\sigma_{ij}(m_{t\bar t},\mu_F)\;,
\end{eqnarray}
where $i,j=g,q,\bar q$, $s$ is the squared proton-proton center-of-mass (c.m.) energy,
${\cal L}_{ij}$ denotes the parton luminosity, and $\hat\sigma_{ij}$ is the corresponding partonic cross section with $\mu_F$ being the factorization scale.
Near threshold, the singular part of the partonic cross section factorizes into a short-distance production coefficient $H$ and a long-distance non-relativistic Green's function $G$,
\begin{eqnarray}
\hat\sigma_{ij}^{\rm thr}(m_{t\bar t})
=
\sum_{R,S}
H_{ij}^{R,S}(m_t)
{\rm Im}\Big\{
G_{R}^{S}({\bf 0},{\bf 0};E+i\Gamma_t) \Big\}
+
\hat\sigma_{ij}^{\rm reg}(m_{t\bar t})\;,
\end{eqnarray}
where $R$ labels the color representation of the $t\bar t$ pair, $S$ denotes the spin channel, and $\hat\sigma_{ij}^{\rm reg}$ contains terms that are regular in the limit $v\to0$.
Under non-relativistic approximation, the spin-dependent contribution is sub-leading order, and hence can be neglected.
The Green's function satisfies the Schr\"{o}dinger equation
\begin{eqnarray}
\left[
-{\nabla^2\over m_t}
+V_R(r)
-\left(E+i\Gamma_t\right)
\right]
G_R({\bf r},{\bf r'};E+i\Gamma_t)
=
\delta^{(3)}({\bf r}-{\bf r'})\;,
\end{eqnarray}
where $V_R(r)$ denotes the static QCD potential approximated by a Coulomb form.
At leading order in the non-relativistic expansion, the Coulomb potentials in the color-singlet and color-octet channels are respectively
\begin{eqnarray}
V_1(r)=-{C_F\alpha_s\over r}\;,
\qquad
V_8(r)=\left({C_A\over2}-C_F\right){\alpha_s\over r}
={\alpha_s\over 2N_c r}\;.
\end{eqnarray}
The color-singlet potential is attractive, whereas the color-octet potential is repulsive.
This distinction is essential in toponium production. Although a near-threshold $t\bar t$ pair can be produced in both color-singlet and color-octet configurations, only the color-singlet channel develops the Coulomb enhancement associated with toponium formation. The physical line shape of toponium production is governed by the Green's function evaluated at the origin
${\rm Im}G_1({\bf 0},{\bf 0};E+i\Gamma_t)$, which resums the Coulombic interactions of the slowly moving $t\bar t$ pair~\cite{Fadin:1990wx,Strassler:1990nw,Fadin:1991zw,Sumino:1992ai}.

In practical collider simulations, it is more convenient to use Green functions in momentum space~\cite{Jezabek:1992np}. In this approach,
the enhancement effect can be implemented by reweighting the relevant color-singlet $S$-wave matrix elements $\calm$ by the ratio of the interacting NRQCD Green's function and the free Green's function~\cite{Fuks:2024yjj,Fuks:2025wtq,Garzelli:2026ctb}
\begin{equation}
|\mathcal{M}|^2 \rightarrow\left|\mathcal{M} \frac{\widetilde{G}\left(E ; p^*\right)}{\widetilde{G}_0\left(E ; p^*\right)}\right|^2,
\end{equation}
where $\widetilde{G}_0$ is the free Green' s function:
$\widetilde{G}_0\left(E ; p^*\right)=\left(E+i \Gamma_t-p^{*2}/m_t\right)^{-1}$,
with $p^*$ being momentum of the top-quark in the toponium rest frame, and $m_t$ and $\Gamma_t$ being the mass and width of top quark, respectively. The interacting Green's function $\widetilde{G}_0\left(E ; p^*\right)$ can be obtained either by Fourier transformation or by
solving the Lippmann-Schwinger equation with the QCD potential~\cite{Jezabek:1992np}.
The threshold prediction may also be combined with soft and collinear threshold resummation for initial state radiation, which is necessary for a realistic prediction of $d\sigma/dm_{t\bar t}$ at the LHC~\cite{Garzelli:2026ctb}.

The dominant toponium contribution at the LHC is generated by gluon fusion.
The leading-order partonic process then becomes
\begin{eqnarray}
g g \to \eta_t \to t\bar t\;,
\end{eqnarray}
where $\eta_t$ denotes the color-singlet, spin-singlet, $S$-wave configuration as
\begin{eqnarray}
\eta_t:\qquad n\;,~~{}^{2S+1}L_J^{[R]}=1\;,~~{}^1S_0^{[1]}
\end{eqnarray}
with $n$ being its principal quantum number and $R$ denoting the $SU(3)_c$ representation.
It gains the quantum numbers of a pseudoscalar $t\bar t$ quasi-bound state.
This is the channel most directly associated with the near-threshold excess reported by CMS and ATLAS~\cite{CMS:2025kzt,ATLAS:2026nrx}.
By contrast, the quark-antiquark annihilation via a gluon in $s$-channel mainly produces a color-octet state $^3S_1^{[8]}$. For this state $^3S_1^{[8]}$ and the color-octet state from gluon fusion $^1S_0^{[8]}$, the QCD potential is repulsive and no analogous bound state enhancement is expected.
A color-singlet and spin-triplet state can also be produced through electroweak $\gamma/Z$ exchange in quark-antiquark annihilation,
\begin{eqnarray}
q\bar q\to \gamma^\ast/Z^\ast \to \psi_t\;,
\qquad
\psi_t:\quad 1\;,~~{}^3S_1^{[1]}\;.
\end{eqnarray}
This contribution is suppressed by the smaller quark-antiquark luminosity at the LHC and by electroweak couplings.
The productions of higher partial waves are further suppressed by powers of the relative velocity $v$. For instance, $P$-wave production involves derivatives of the wave function at the origin and is parametrically smaller than the leading $S$-wave contribution.
For the purpose of describing the dominant threshold enhancement, it is therefore well-motivated to focus on the color-singlet $^1S_0$ channel produced through gluon fusion.

Since the chromo-electric dipole operator does not introduce any additional color factors, the color flow remains the same as the SM case for all channels under consideration. For the quark-pair annihilation channel, the color flow is trivial.
For the gluon fusion channel, there are three SM diagrams and the corresponding total amplitude with the explicit color factors can be written as
\bee
\calm
=
i f^{ABC} (T^C)^{a}_{~b} (\calm_{1, AB})^{a}_{~b}
+
(T^A T^B)^{a}_{~b} (\calm_{2, AB})^{a}_{~b}
+
(T^B T^A)^{a}_{~b} (\calm_{3, AB})^{a}_{~b} \,,
\ene
where $f^{ABC}$ are the antisymmetric structure constants of the $SU(3)_c$ group. After a straightforward calculation, the total squared amplitude can be written as
\bee
\Big| \calm \Big|^2
=
\sum_{A,B} \sum_{a,b}
\rho_{ij}
\big(\calc_{ij}^{BAAB} \big)^{aa}_{\;\;\;\;bb}
=
\rho_{ij} \calc_{ij} \,,
\quad
\calc_{ij}
=
\sum_{A,B} \sum_{a,b}
\big(\calc_{ij}^{BAAB} \big)^{aa}_{\;\;\;\;bb}  \,,
\ene
where the reduced density matrix and the color matrix are defined as
\bea
\rho_{ij} &=& \calm_{i,AB} \calm_{j,AB}^\dag\,,
\\[1em]
\calc_{ij}
&=&
\sum_{a,b}\sum_{A,B}
\left(
\begin{array}{cc}
  \big(T^B T^A \big)^{~a}_{b} \big(T^A T^B\big)^{a}_{~b}
  \;&\;
  \big( T^B T^A  \big)^{~a}_{b} \big( T^B T^A\big)^{a}_{~b}
  \\[1em]
   \big( T^A T^B  \big)^{~a}_{b} \big( T^A T^B\big)^{a}_{~b}
   \;&\;
   \big(T^A T^B  \big)^{~a}_{b} \big( T^B T^A \big)^{a}_{~b}
\end{array}
\right)\,.
\ena
The explicit expressions for the color matrices in the singlet and octet cases are respectively given by
\bee
\calc_{ij}^{[1]}
=
\frac{N_C^2 -1 }{4N_C}
\left(
\begin{array}{cc}
   1
  \;&\;
  1
  \\[1em]
   1
   \;&\;
   1
\end{array}
\right)\;,
\quad
\calc_{ij}^{[8]}
=
\frac{1}{4N_C}
\left(
\begin{array}{cc}
  (N_C^2-1)^2
  \;&\;
  1 - N_C^2
  \\[1em]
   1 - N_C^2
   \;&\;
   (N_C^2-1)^2
\end{array}
\right)\;,
\ene
where the number of color types $N_C=3$. We accordingly modify the color matrix elements in MagGraph to implement the production of color-singlet or color-octet toponium.

In this work, we describe toponium production through the universal color-singlet $S$-wave threshold enhancement of the $t\bar t$ production amplitude.
The short-distance partonic process fixes the hard production coefficient and the spin/color quantum numbers of the heavy quark pair. The non-relativistic QCD dynamics is encoded in the Green's function or, equivalently, in the toponium line shape.
This separation allows the resonant threshold contribution to be consistently combined with the continuum $t\bar t$ amplitude and with the subsequent decay spin correlations of the top and anti-top quarks.
It also provides a natural framework for studying anomalous interactions near threshold. The new short-distance operators modify the hard production vertices, whereas the long-distance QCD enhancement remains governed by the same color-singlet Green's function.
Therefore, the toponium production at the LHC is not only a manifestation of QCD dynamics in the heaviest quark system, but also a sensitive laboratory for probing the spin and CP structure of top quark interactions near the $t\bar t$ kinematic threshold.

\begin{figure}[htb]
\begin{center}
\includegraphics[width=0.48\textwidth]{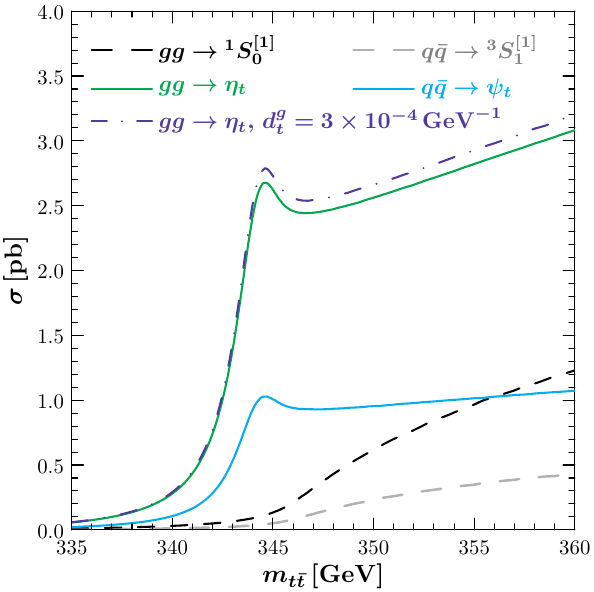}
\,
\includegraphics[width=0.48\textwidth]{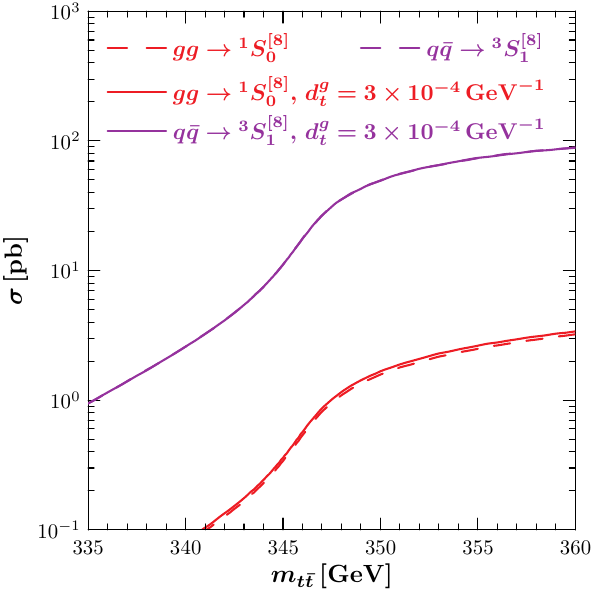}
\caption{\it
Invariant-mass dependence of the parton-level cross sections for the different $t\bar t$ production channels near threshold.
The left panel shows the color-singlet contributions, including the continuum channels $gg\to{}^1S_0^{[1]}$ (black dashed line) and $q\bar q\to{}^3S_1^{[1]}$ (gray dashed line), together with the corresponding toponia $\eta_t$ (green solid line) and $\psi_t$ (blue solid line) production cross sections, for which Coulomb-like threshold corrections are included. The right panel shows the continuum color-octet channels, $gg\to{}^1S_0^{[8]}$ (red dashed line) and $q\bar q\to{}^3S_1^{[8]}$ (purple dashed line). The SM predictions are compared with those obtained for a nonzero top quark CEDM coupling, $d_t^g=3\times10^{-4}~{\rm GeV}^{-1}$ (dashed-dotted line in the left panel and solid lines in the right panel).
}
\label{fig:cc:ced}
\end{center}
\end{figure}
In Fig.~\ref{fig:cc:ced}, we show the parton-level cross sections for color-singlet (left panel) and color-octet (right panel) contributions to near-threshold $t\bar t$ production at the LHC, as a function of invariant mass $m_{t\bar{t}}$ being the c.m. energy at parton-level. In the color-singlet channel, as shown by the solid lines in the left panel, the attractive QCD potential leads to a pronounced threshold enhancement. The pseudoscalar spin-singlet state $\eta_t$, corresponding to the $^1S_0^{[1]}$ configuration produced predominantly through gluon fusion, provides the dominant contribution in the threshold region. The spin-triplet color-singlet state $\psi_t$ is produced through the electroweak process $q\bar q\to\gamma^\ast/Z^\ast\to t\bar t$ and is likewise enhanced by the attractive color-singlet potential. Its contribution remains a few times smaller than that of $\eta_t$. The corresponding continuum color-singlet contributions from $gg\to t\bar t$ and $q\bar q\to\gamma^\ast/Z^\ast\to t\bar t$ (dashed lines) are comparatively suppressed near threshold. In contrast, as shown in the right panel, the color-octet contributions of $^1S_0^{[8]}$ and $^3S_1^{[8]}$ do not exhibit a toponium-like threshold enhancement, owing to the repulsive nature of the color-octet QCD potential.

Fig.~\ref{fig:cc:ced} also illustrates the effect of a nonzero top quark CEDM by taking $d_t^g=3\times10^{-4}~{\rm GeV}^{-1}$ as a benchmark. For this value, as shown by the dash-dotted line in the left panel, the CEDM induces a sizable modification of the color-singlet $\eta_t$ line shape. Its effect on the color-octet contributions is nearly indistinguishable from the $d_t^g=0$ prediction, as seen by the solid lines in the right panel. These results indicate that the near-threshold invariant-mass spectrum alone has limited sensitivity to the CEDM. Far greater sensitivity can be instead afforded by CP-odd spin correlations and asymmetry observables, in which interference between the SM and CEDM-induced amplitudes generates contributions linear in $d_t^g$.

\begin{table}[thb]
\renewcommand\arraystretch{1.6}
\begin{center}
\caption{\it
Cross sections for near-threshold $t\bar t$ production with invariant mass window $330~{\rm GeV}<m_{t\bar{t}}<360~{\rm GeV}$ at LHC13 and LHC14, separated according to the initial state and the color/spin configuration of the $t\bar t$ system.
Results are shown for the SM, $d_t^g=0$, and for the benchmark CEDM coupling $d_t^g=3\times10^{-4}~{\rm GeV}^{-1}$.
For the $gg$ initial state, the color-singlet $\eta_t$ (${}^1S_0^{[1]}$) and color-octet ${}^1S_0^{[8]}$ contributions are shown. For the $q\bar q$ initial state, the color-singlet $\psi_t$ (${}^3S_1^{[1]}$) contribution mediated by $\gamma^\ast/Z^\ast$ exchange and the color-octet ${}^3S_1^{[8]}$ contribution mediated by gluon exchange are shown. Cross sections are given in unit of pb.
}
\begin{tabular}{c c c c c c}
\\\hline\hline
\multirow{2}{*}{ {\rm Channels}  }  & \multirow{2}{*}{ {\rm Color State}  }
& \multicolumn{2}{c}{ $d_t^g=0\, [{\rm GeV}^{-1}]$}
& \multicolumn{2}{c}{ $d_t^g=3\times10^{-4}\, [{\rm GeV}^{-1}]$} \hspace{2em}
\\ \cline{3-6}
&  & LHC13 & LHC14  & LHC13 & LHC14
\\\hline
\multirow{2}{*}{ $gg\to t\bar{t}$ }
& $\eta_t$ &  $13.65$ pb &  $15.99$ pb &  $14.20$ pb  &  $16.64$ pb
\\
& ${}^1S_{0}^{[8]}$  &  $8.50$ pb  &  $9.95$ pb  &  $8.96$ pb  &  $10.51$ pb
\\\hline
\multirow{2}{*}{ $q\bar{q}\to t\bar{t}$ }
& $\psi_t$ &  $0.0543$ pb  &  $0.0602$ pb &  --  &  --
\\
& ${}^3S_{1}^{[8]}$  &  $3.01$ pb  &  $3.33$ pb  &  $3.01$ pb  &  $3.33$ pb
\\\hline\hline
\end{tabular}
\label{tab:ilhc}
\end{center}
\end{table}
Tab.~\ref{tab:ilhc} summarizes the contributions of different initial states and color configurations to the total cross sections of $t\bar t$ production near threshold at the LHC with $\sqrt{s}=13$ TeV (LHC13) and $\sqrt{s}=14$ TeV (LHC14). They are obtained by integrating over the invariant mass window $330~{\rm GeV}<m_{t\bar{t}}<360~{\rm GeV}$.
Given the convolution with the parton distribution functions, one can see that the gluon-fusion channels provide the dominant contributions. For $d_t^g=0$, the color-singlet $\eta_t$ contribution from gluon fusion is $13.65$ pb at LHC13 and $15.99$ pb at LHC14. The corresponding color-octet ${}^1S_0^{[8]}$ contribution is $8.50$ pb and $9.95$ pb, respectively. By comparison, the $q\bar q$-initiated contributions are substantially smaller.
The suppression is particularly pronounced for the color-singlet $\psi_t$ channel produced through $q\bar q\to\gamma^\ast/Z^\ast\to t\bar t$. Its cross section is only $0.0543$ pb at LHC13 and $0.0602$ pb at LHC14. These values are smaller than the corresponding color-singlet gluon-fusion contribution by a factor of approximately $250$--$270$. This strong suppression can be attributed to the smaller $q\bar q$ parton luminosity in the relevant threshold region, together with the electroweak nature of the $\gamma/Z$-mediated production process. The contribution of color-octet ${}^3S_1^{[8]}$ from $q\bar q\to g^\ast\to t\bar t$ is considerably larger, reaching $3.01$ pb at LHC13 and $3.33$ pb at LHC14, but remains below the corresponding gluon-fusion contribution.

For the benchmark CEDM coupling $d_t^g=3\times10^{-4}~{\rm GeV}^{-1}$, the gluon-fusion cross sections receive modest enhancements. The $\eta_t$ contribution increases by about $4\%$, while the color-octet ${}^1S_0^{[8]}$ contribution approximately increases by $5$--$6\%$. In contrast, the effect on the $q\bar q$-initiated color-octet channel is negligible at the precision shown in the table. The color-singlet $q\bar q\to\gamma^\ast/Z^\ast\to t\bar t$ channel receives no direct CEDM correction at this order, since the CEDM modifies the top-gluon interaction rather than the electroweak $t\bar t\gamma$ and $t\bar t Z$ vertices. Given its strong suppression relative to the gluon-fusion contribution, we neglect the $\gamma/Z$-mediated color-singlet $q\bar q$ channel in the subsequent analysis and focus on the dominant gluon-fusion contribution.

\begin{figure}[htb]
\centering
\includegraphics[width=0.98\textwidth]{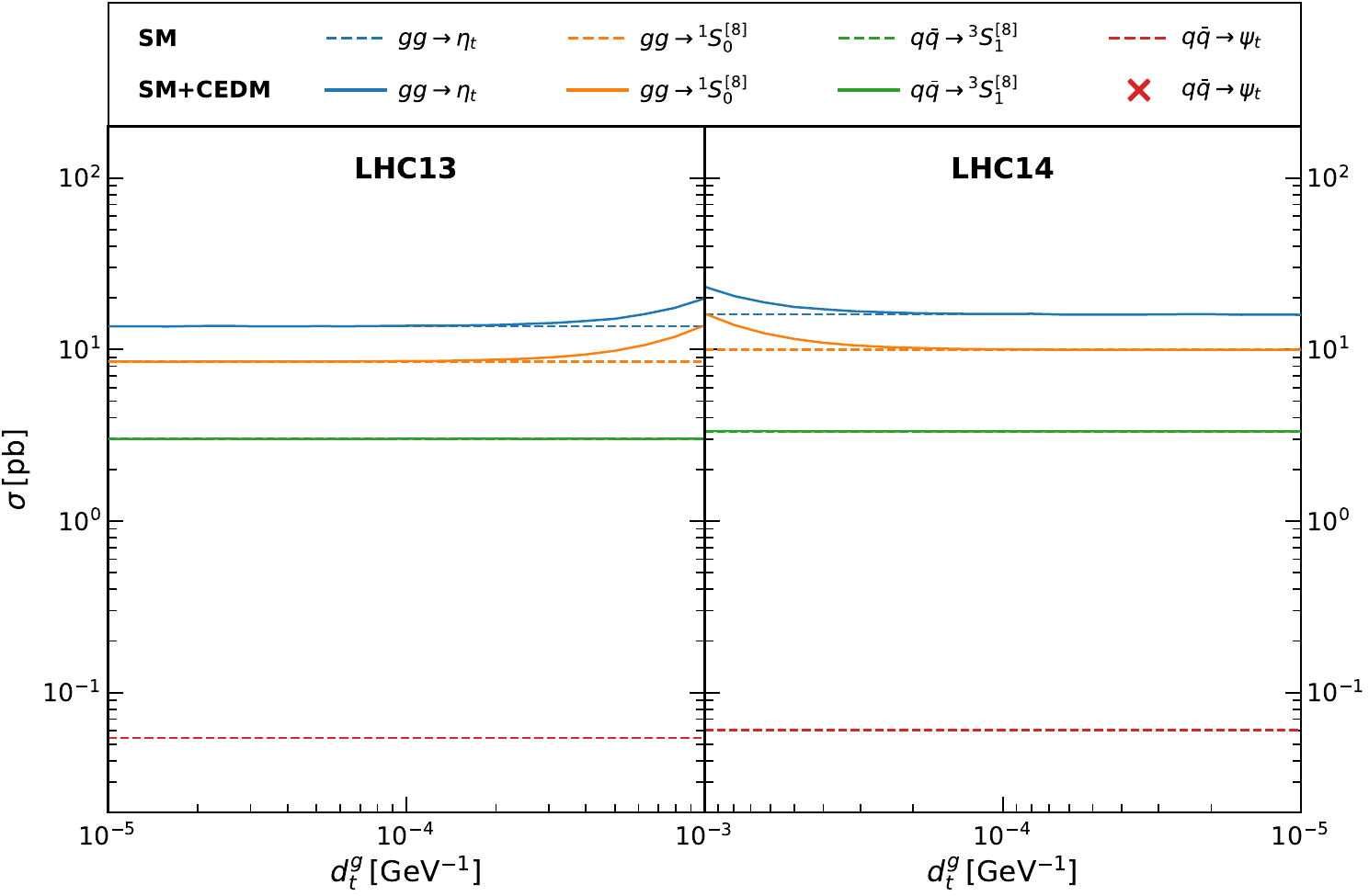}
\caption{\it
Dependence of the near-threshold $t\bar t$ production cross sections on the top quark CEDM coupling $d_t^g$ at LHC13 (left) and LHC14 (right).
The cross sections are obtained by integrating over the invariant mass window $330~{\rm GeV}<m_{t\bar{t}}<360~{\rm GeV}$.
The blue, orange, green, and red curves correspond to the $gg\to\eta_t$, $gg\to{}^1S_0^{[8]}$, $q\bar q\to{}^3S_1^{[8]}$, and $q\bar q\to\psi_t$ channels, respectively.
Dashed curves denote the SM predictions, while solid curves show the corresponding results including the CEDM contribution.
}
\label{fig:xsd}
\end{figure}
Fig.~\ref{fig:xsd} shows the dependence of the near-threshold $t\bar t$ production cross sections on the top quark CEDM coupling $d_t^g$ at LHC13 (left panel) and LHC14 (right panel), after integrating over the invariant mass window \(330~{\rm GeV}<m_{t\bar{t}}<360~{\rm GeV}\). The CEDM dependence is most pronounced in the gluon-fusion channels. For \(d_t^g\lesssim10^{-4}~{\rm GeV}^{-1}\), the cross sections of \(gg\to\eta_t\) and \(gg\to{}^1S_0^{[8]}\) production remain close to their SM predictions, whereas an increasingly visible enhancement develops at larger values of \(d_t^g\). In contrast, the color-octet \(q\bar q\to{}^3S_1^{[8]}\) contribution is essentially unchanged over the considered range. The electroweak color-singlet \(q\bar q\to\psi_t\) channel receives no direct CEDM correction at this order. The same qualitative behavior is observed at LHC13 and LHC14.

The weak sensitivity of the inclusive rates at small \(d_t^g\) can be understood from their CP-even nature in the SM. The interference between the CP-even SM amplitude and the CP-odd CEDM amplitude is absent or strongly suppressed after phase space integration, leaving a predominantly quadratic dependence on \(d_t^g\). As a result, sizable deviations in the inclusive rates arise only for relatively large values of the CEDM. By contrast, suitably constructed CP-odd observables can retain the SM--CEDM interference term and therefore depend linearly on \(d_t^g\). They would provide greater sensitivity to small CP-violating effects. In the next section, we will explore this sensitivity using CP-odd angular asymmetries and spin-correlation observables.

\section{Asymmetries and prospective sensitivity to top quark CEDM}
\label{sec:Res}

\subsection{Asymmetries}
\label{sec:asymmetries}

Because the top quark decays weakly before hadronization, the observed final state in experiment is not a stable toponium resonance but rather
\begin{eqnarray}
pp\to \eta_t+X\to t\bar t+X
\to bW^+ \bar b W^-+X\;.
\end{eqnarray}
The leptonic channel of top quarks then becomes
\begin{eqnarray}
pp\to \eta_t+X\to
b\ell^+\nu_\ell +
\bar b\ell'^-\bar\nu_{\ell'}+X\;.
\end{eqnarray}
This channel is especially useful because charged leptons retain strong information about the spin state of the parent top and anti-top quarks.
The $^1S_0^{[1]}$ toponium contribution has a characteristic spin-singlet structure and therefore modifies not only the invariant-mass spectrum but also spin-correlation observables.
Recent LHC analyses have exploited this feature by studying the near-threshold $m_{t\bar t}$ distribution together with angular observables sensitive to the top-pair spin correlations~\cite{CMS:2025kzt,ATLAS:2026nrx}. We will also analyze the asymmetries based on the leptonic channel of toponium at LHC.

The spin-correlation method provides a direct way to convert the spin information of the produced
$t\bar t$ system into experimentally accessible angular asymmetries of the charged leptons from
top decay.  In the narrow width approximation, the production and decay density matrices can be
factorized, and the spin dependence of the production process is encoded in the normalized
$t\bar t$ spin density matrix. Since the charged lepton from the semi-leptonic top decay is a maximal analyzer of the parent top quark spin, the leptonic angular distribution can be used to study spin correlations of the parent top-pair. In the rest frames of top and anti-top quarks, the leptonic angular distribution can be written
schematically as
\begin{eqnarray}
{1\over \sigma}
{d\sigma\over d\Omega_+d\Omega_-}
=
{1\over 16\pi^2}
\left[
1
+
\kappa_+ B_i^+ \hat n_+^i
+
\kappa_- B_i^- \hat n_-^i
+
\kappa_+\kappa_- C_{ij}\hat n_+^i \hat n_-^j
\right],
\end{eqnarray}
where $\hat n_+$ and $\hat n_-$ denote the directions of $\ell^+$ and $\ell^-$ in the rest frames of top and anti-top quark, respectively, and $\kappa_\pm\simeq 1$ are the charged-lepton spin-analyzing powers.
As pointed out in Ref.~\cite{Ma:2017vve}, the reconstruction of two separate reference frames can be avoided by expressing the spin correlations in a single reference frame using a helicity-projected basis. The corresponding density-matrix elements can then be projected out through a complete set of sign-weighted asymmetries in Eq.~(\ref{eq:asym}).
The $\xi$ there denotes an appropriate function of the polar and azimuthal angles of the charged leptons. A key advantage of this construction is that the same formalism can be applied in any chosen reference frame, such as the laboratory frame, the $t\bar t$ c.m. frame, or a frame adapted to the hard-scattering process. This flexibility allows the reference frame to be optimized for sensitivity to the relevant spin-density-matrix elements while simultaneously reducing uncertainties associated with event reconstruction~\cite{Ma:2017vve,Ma:2018ott}.

This method is particularly well suitable for studying CP violation in toponium production near the \(t\bar t\) threshold. In this region, the non-relativistic QCD dynamics governs the threshold line shape, while the short-distance production interaction determines the spin structure of the \(t\bar t\) system. In practice, however, experimental uncertainties limit the precision with which the threshold line shape can be reconstructed. Consequently, small distortions of the invariant mass spectrum induced by a nonzero top quark CEDM are difficult to resolve. This motivates the use of spin-correlation observables, which provide complementary and potentially more sensitive probes of CP-violating effects. In particular, for a CP-odd observable, the corresponding asymmetry vanishes in the CP-conserving limit, up to possible small contributions from the SM and backgrounds, whereas a nonzero CEDM can induce a nonvanishing asymmetry.

The total cross section is a CP-even observable and is therefore predominantly sensitive to contributions quadratic in the CEDM after phase space integration. By contrast, CP-odd lepton angular asymmetries can directly probe the interference between the SM and CEDM-induced amplitudes
\begin{eqnarray}
d\sigma_{\rm CP\mbox{-}odd}
\propto
2\,{\rm Re}
\left[
{\cal M}_{\rm toponium}
{\cal M}_{\rm CEDM}^{\ast}
\right]
\sim {\cal O}(d_t^g)\;,
\end{eqnarray}
which is linearly sensitive to the CEDM coupling.
It is also evident that the interference term is enhanced by the near-threshold QCD effect.
In this work, we adopt this strategy to construct CP-even and CP-odd asymmetry observables in the near-threshold region and use them to probe CP violation in toponium production at the LHC.
They include
\begin{eqnarray}
&&\xi_1=\cos\theta_{\ell^+}\;,~~~\xi_2=\cos\theta_{\ell^-}\;,~~~\xi_3=\cos\theta_{\ell^-}\cos\theta_{\ell^+}\;,~~~\xi_4=\cos\phi_{\ell^+}\;,\nonumber\\
&&\xi_5=\cos\phi_{\ell^-}\;,~~~\xi_6=\cos(\phi_{\ell^-}-\phi_{\ell^+})\;,~~~\xi_7=\cos(\phi_{\ell^-}+\phi_{\ell^+})\;,~~~ \xi_8=\cos\phi_{\ell^-}\cos\theta_{\ell^+}\;,\nonumber\\
&&\xi_9=\cos\theta_{\ell^-}\cos\phi_{\ell^+}\;,~~~\xi_{10}=\sin\phi_{\ell^+}\;,~~~\xi_{11}=\sin\phi_{\ell^-}\;,~~~\xi_{12}=\sin(\phi_{\ell^-}+\phi_{\ell^+})\;,\nonumber\\
&&\xi_{13}=\sin(\phi_{\ell^-}-\phi_{\ell^+})\;,~~~\xi_{14}=\cos\theta_{\ell^-}\sin\phi_{\ell^+}\;,~~~\xi_{15}=\cos\theta_{\ell^+}\sin\phi_{\ell^-}\;,
\end{eqnarray}
where $\xi_{1-9}$ are CP-even and $\xi_{10-15}$ are CP-odd.
All observables are defined in the \(t\bar t\) rest frame. The \(z\)-axis is chosen along the top quark momentum, while the \(x\)-axis is defined by the component of the total missing transverse momentum measured in the laboratory frame. The \(y\)-axis is then fixed by \(\hat{\boldsymbol y}=\hat{\boldsymbol z}\times\hat{\boldsymbol x}\), forming a right-handed coordinate system.

\begin{figure}[!ht]
\centering
\includegraphics[width=0.99\textwidth]{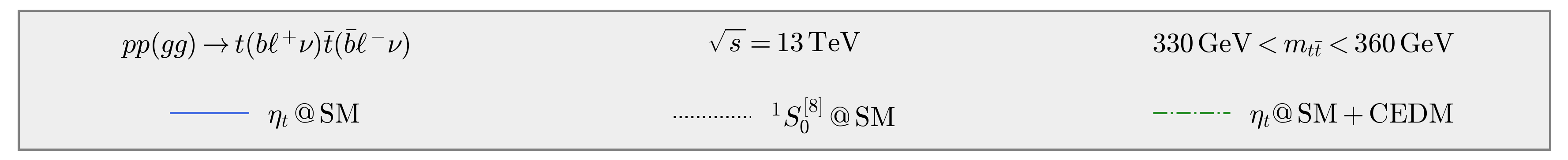}
\\
\includegraphics[width=0.3\textwidth]{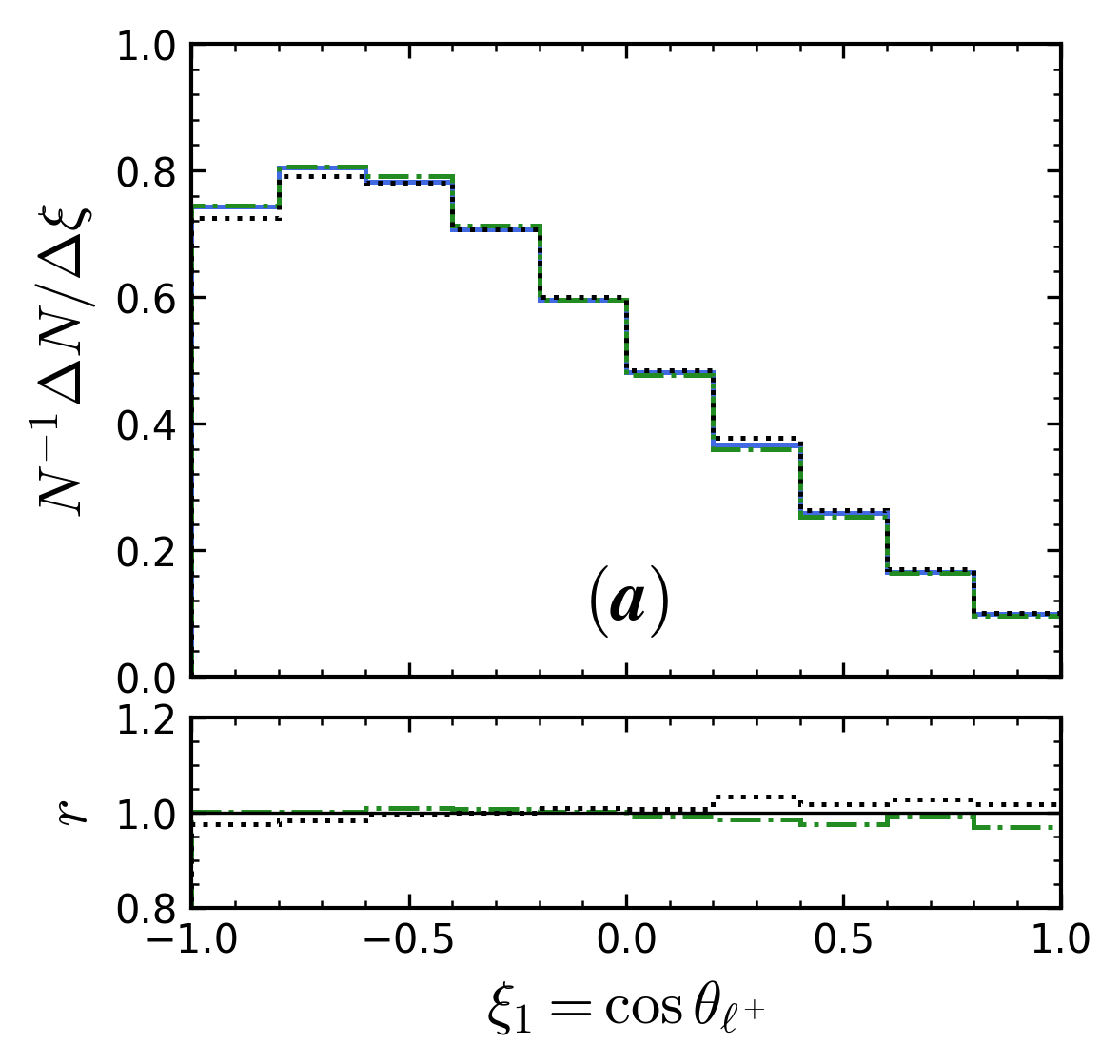}
\hfill
\includegraphics[width=0.3\textwidth]{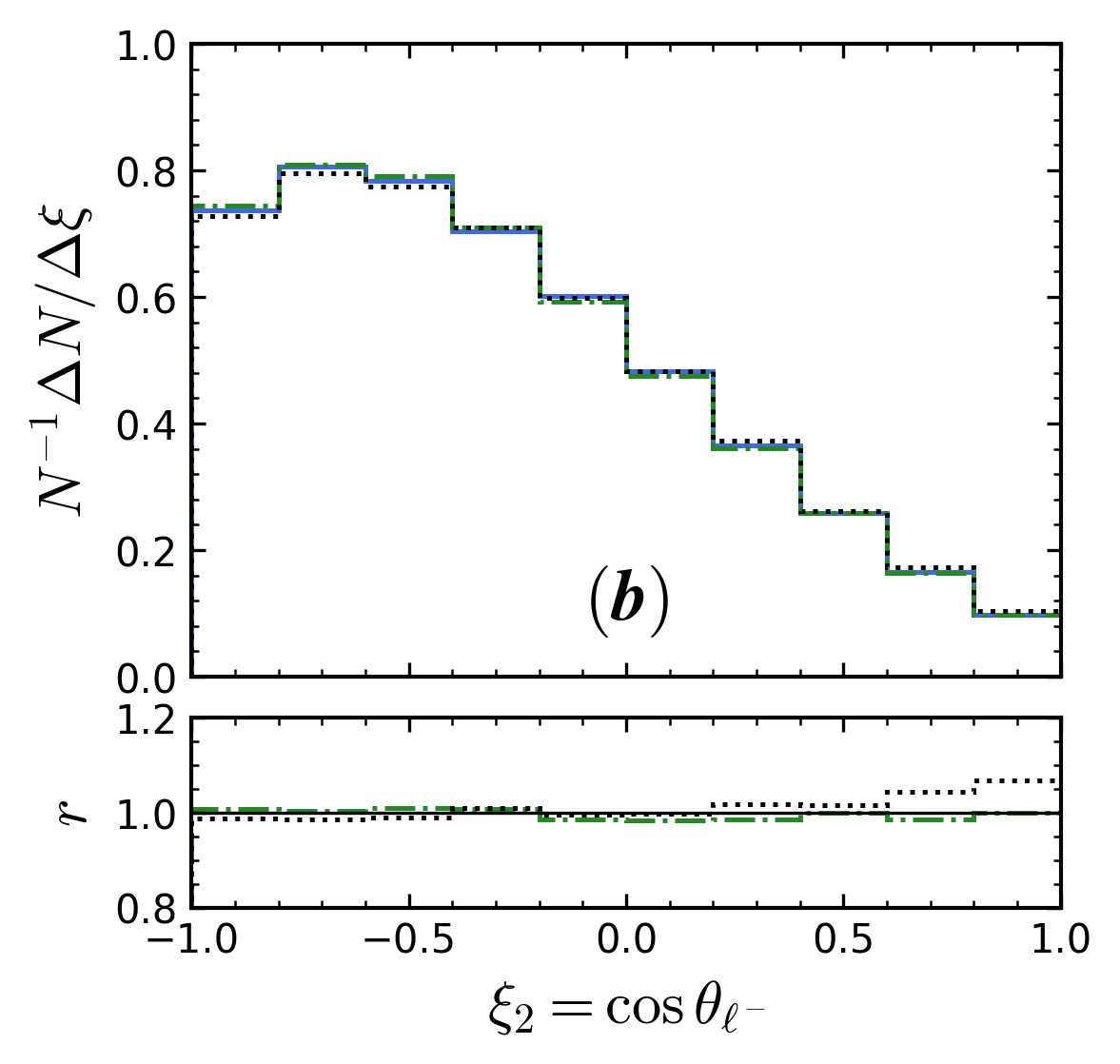}
\hfill
\includegraphics[width=0.3\textwidth]{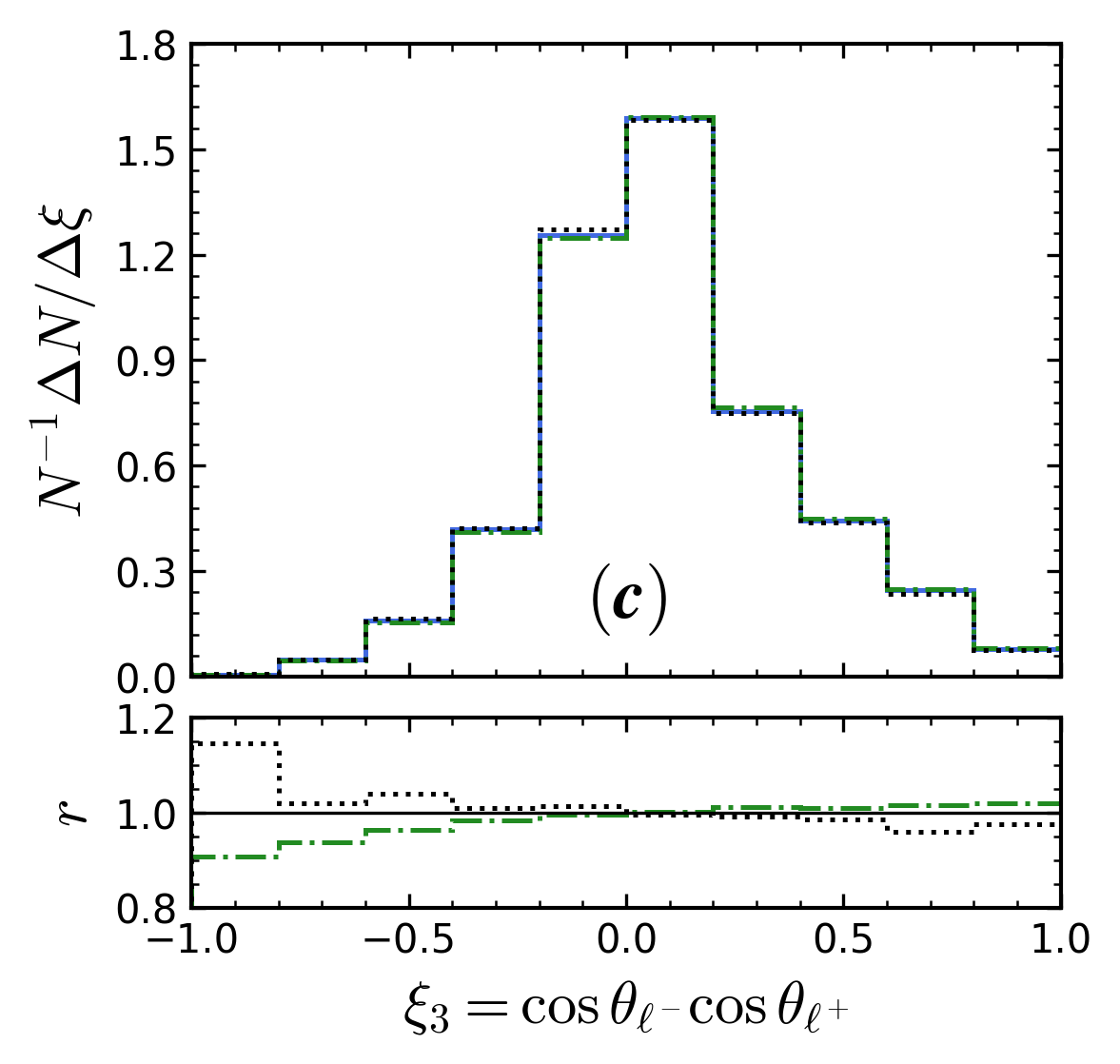}
\hfill
\includegraphics[width=0.3\textwidth]{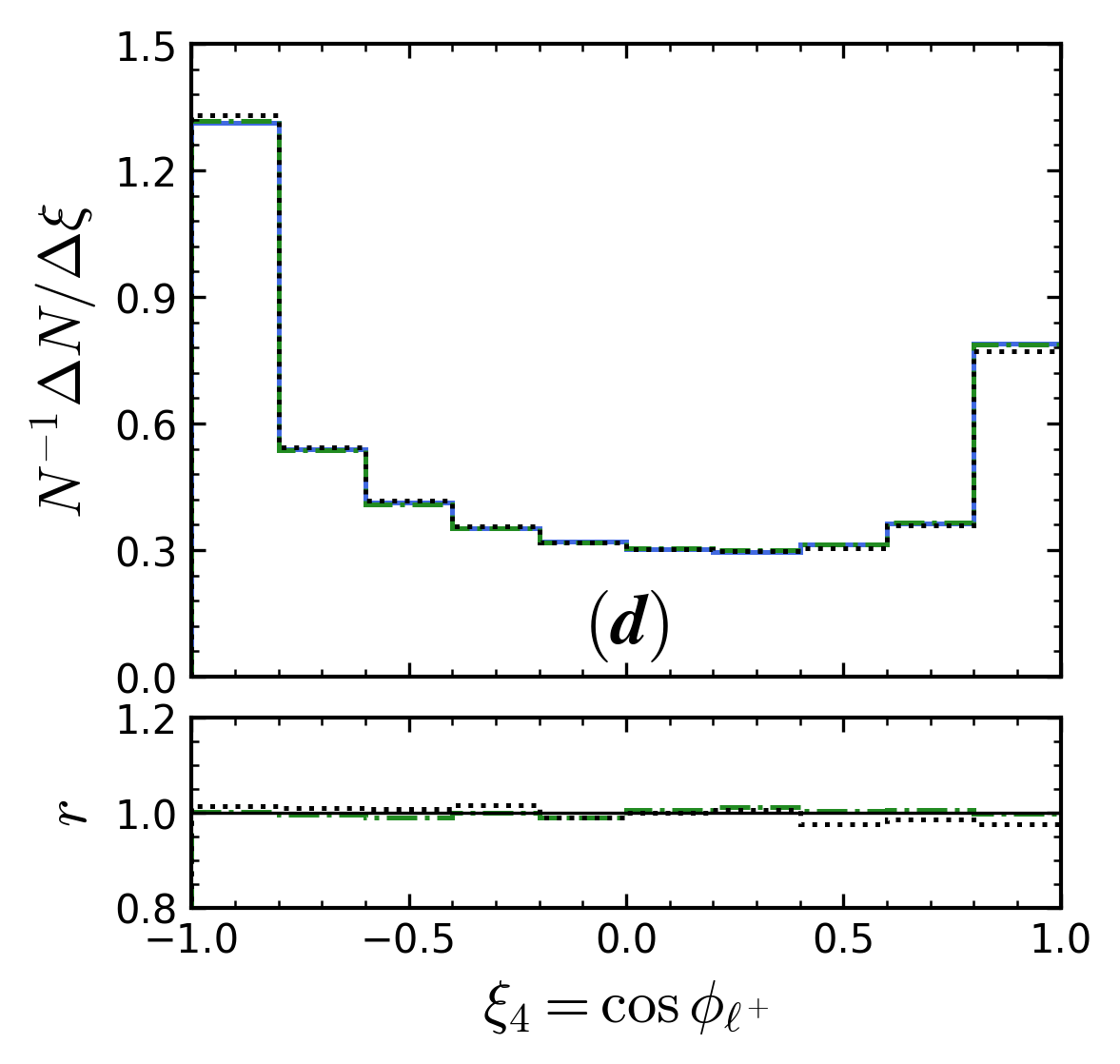}
\hfill
\includegraphics[width=0.3\textwidth]{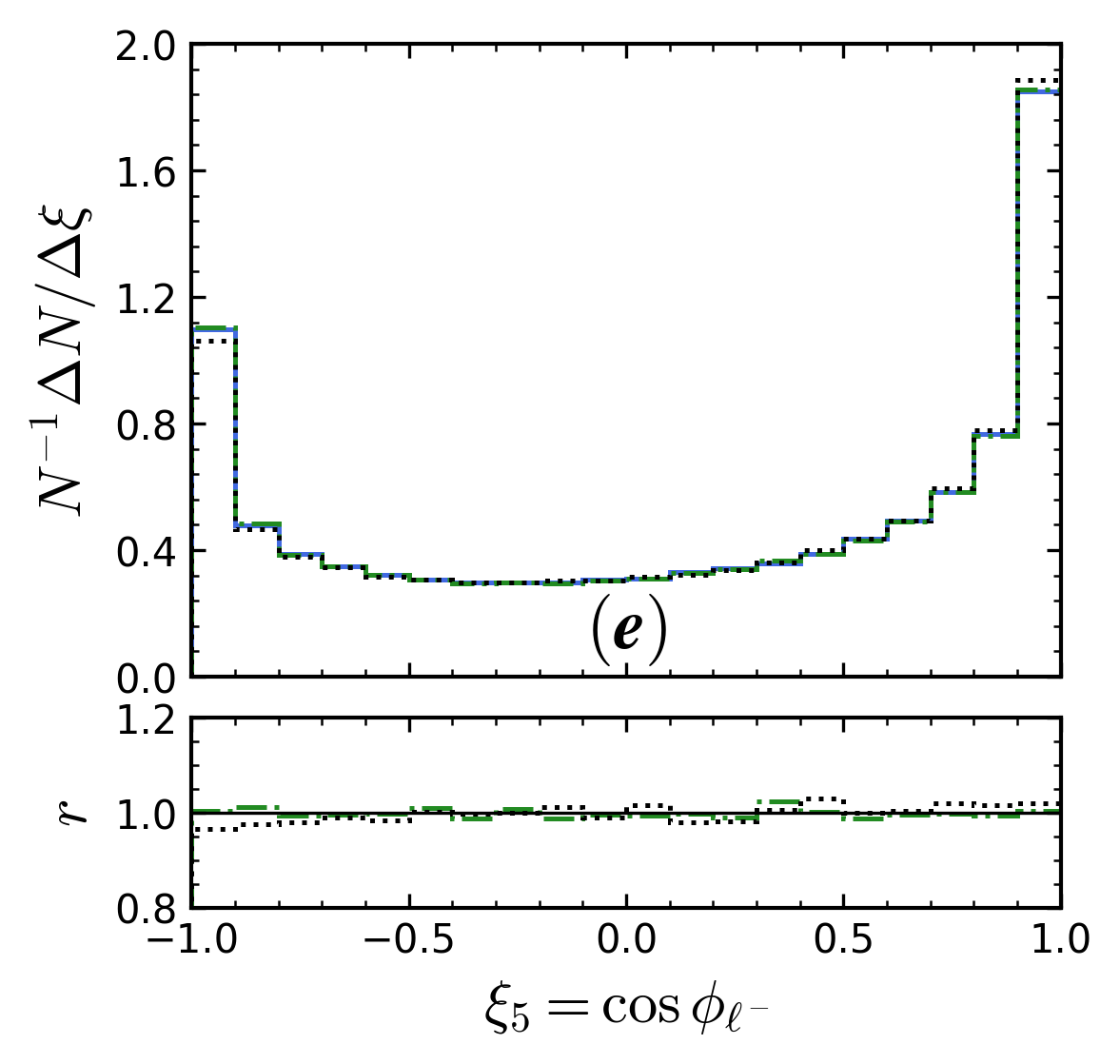}
\hfill
\includegraphics[width=0.3\textwidth]{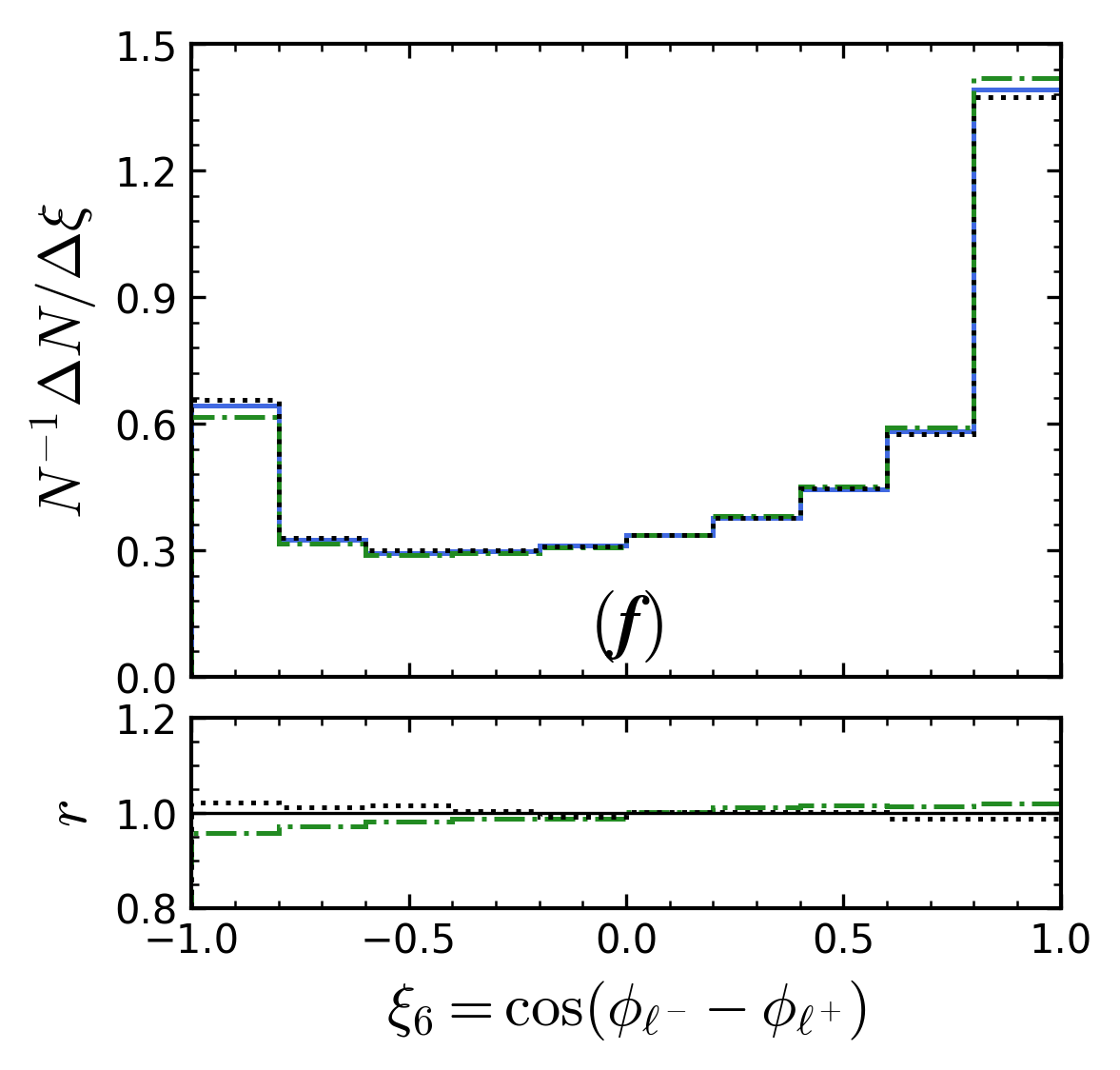}
\hfill
\includegraphics[width=0.3\textwidth]{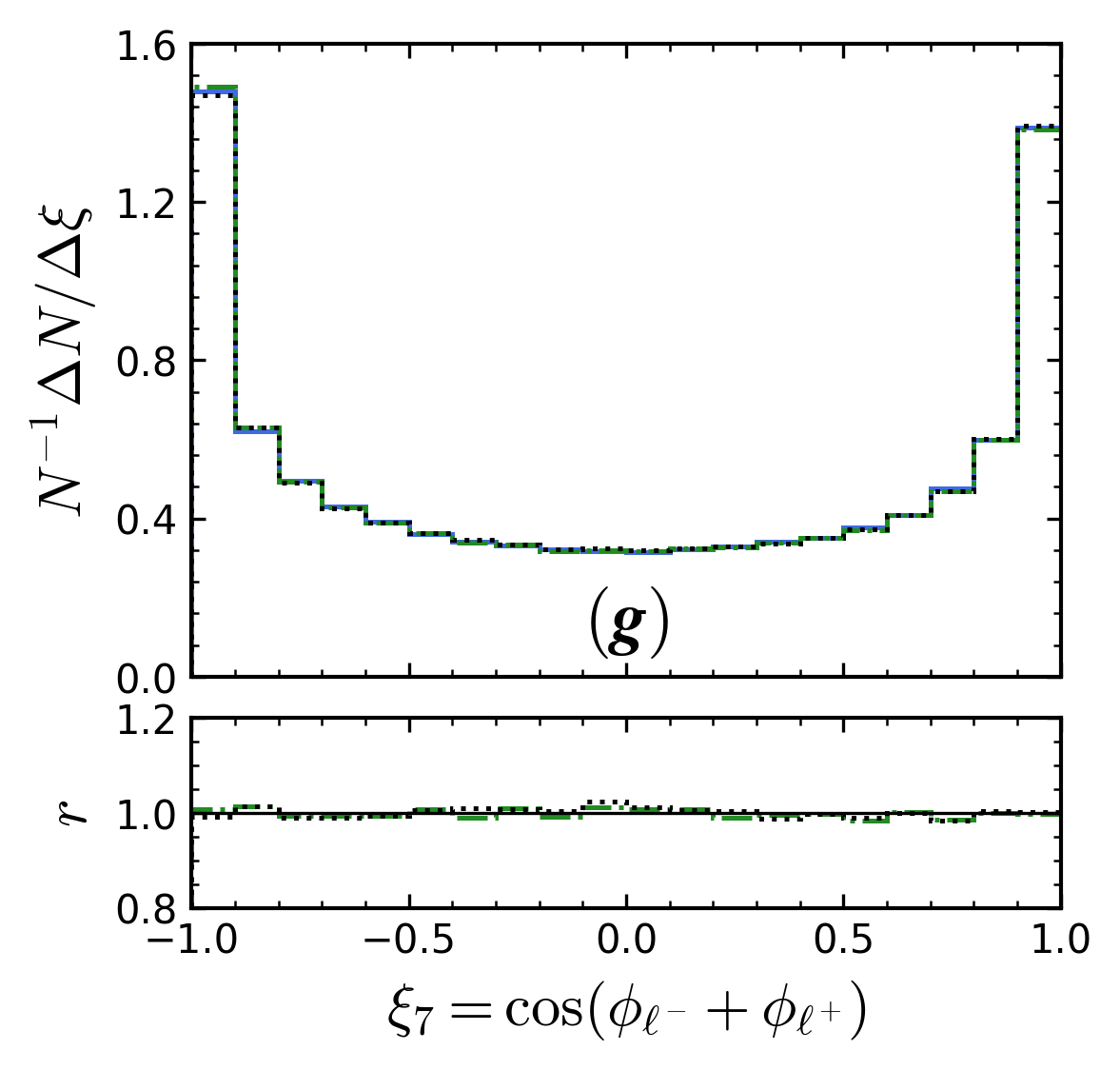}
\hfill
\includegraphics[width=0.3\textwidth]{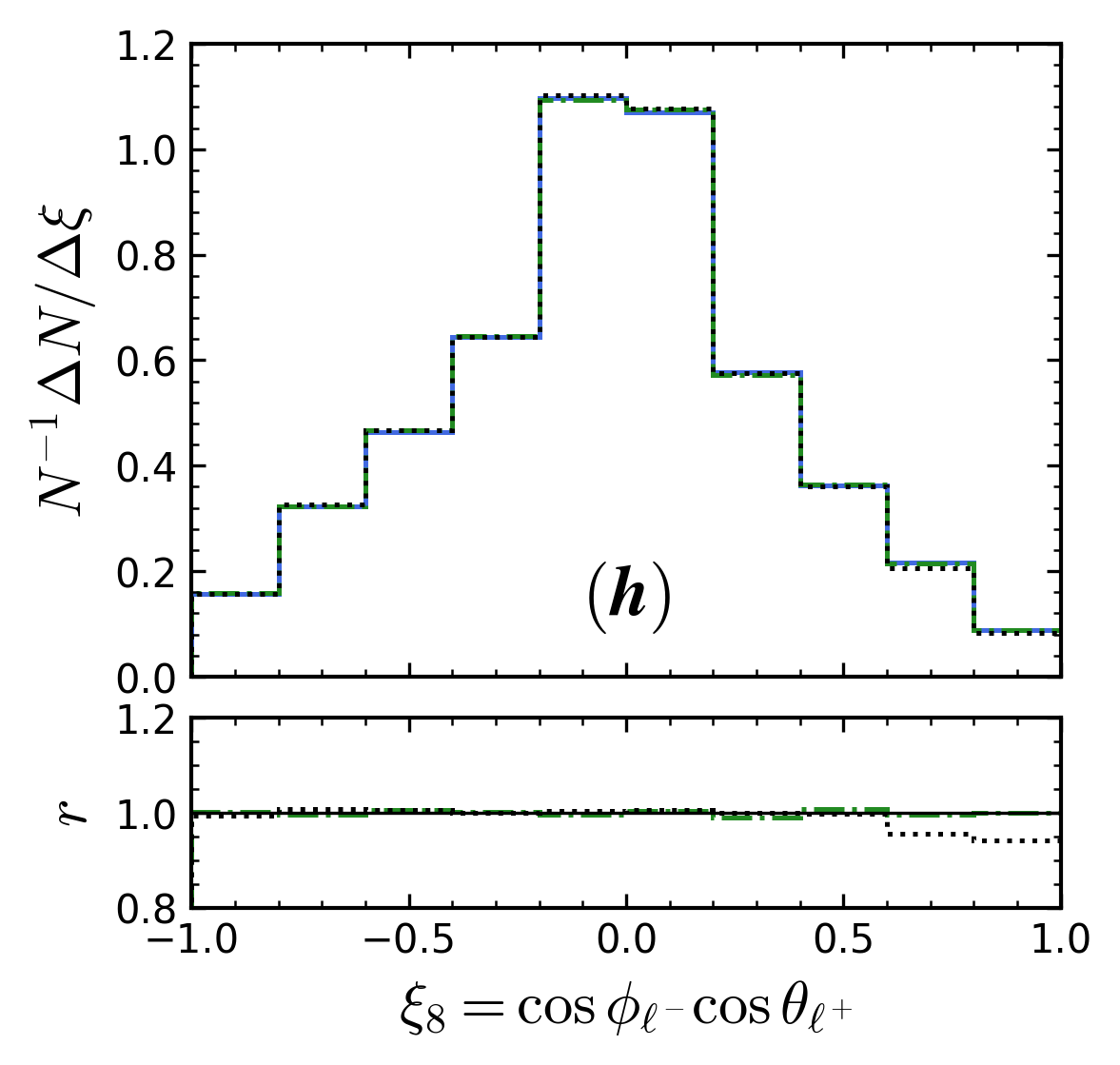}
\hfill
\includegraphics[width=0.3\textwidth]{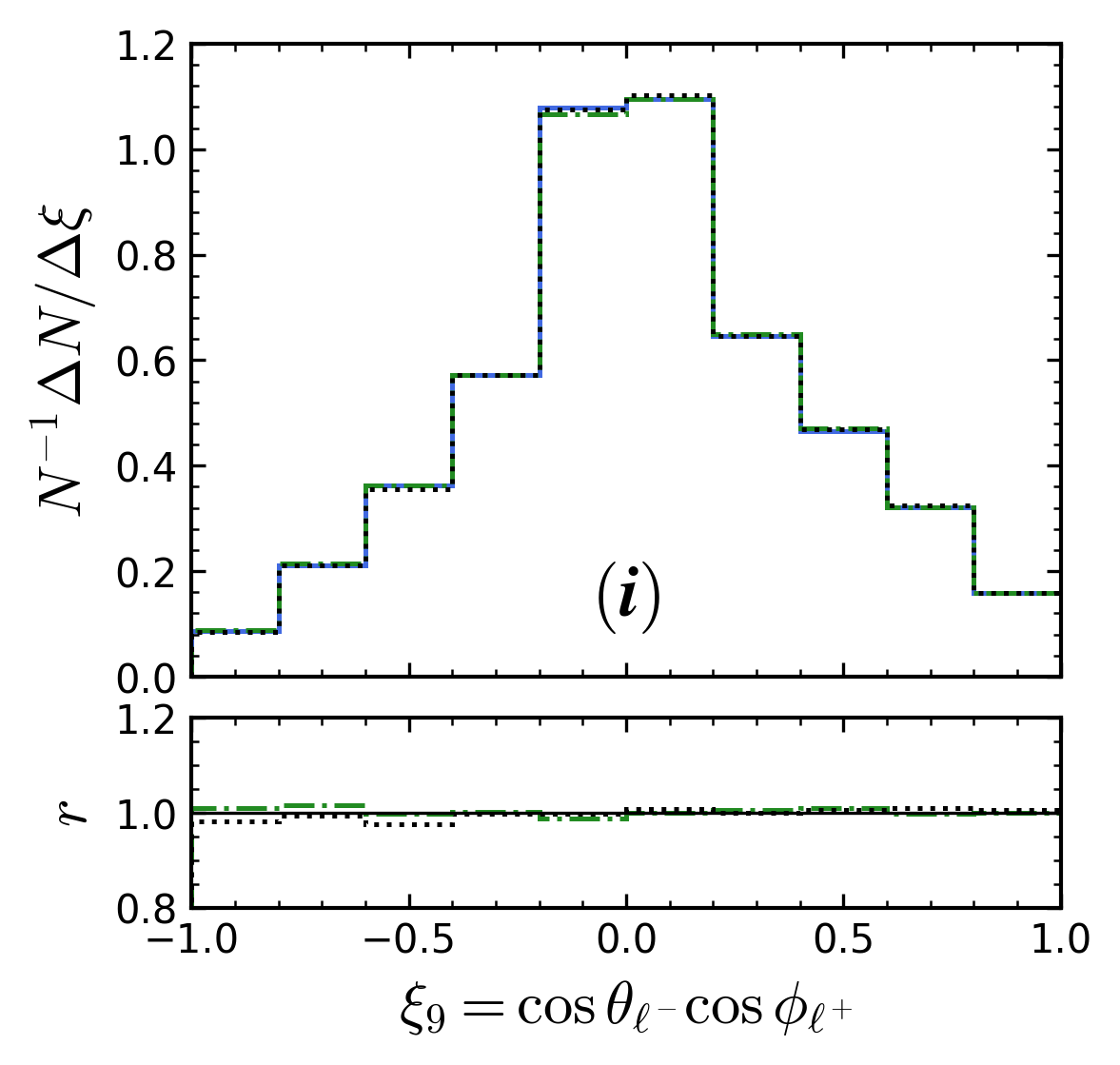}
\caption{\it
Normalized distributions of CP-even lepton angular observables in the dileptonic process
$pp(gg)\to t(\ell^+\nu_\ell b)\bar t(\ell^-\bar\nu_\ell\bar b)$ at
$\sqrt{s}=13~{\rm TeV}$ in the near-threshold region
$330~{\rm GeV}<m_{t\bar t}<360~{\rm GeV}$.
The upper subpanels show the normalized distributions,
$N^{-1}\Delta N/\Delta\xi$, as functions of the angular variable $\xi$,
while the lower subpanels display the corresponding ratios $r$ with respect to
the SM $\eta_t$ prediction.
The blue solid curves correspond to the SM color-singlet $\eta_t$ contribution,
the green dash-dotted curves show the corresponding prediction including the
top quark CEDM, and the black dotted curves denote the color-octet
${}^1S_0^{[8]}$ contribution.
}
\label{fig:asy}
\end{figure}

\begin{figure}[tb]
\centering
\includegraphics[width=0.99\textwidth]{asyLegend.png}
\\
\includegraphics[width=0.31\textwidth]{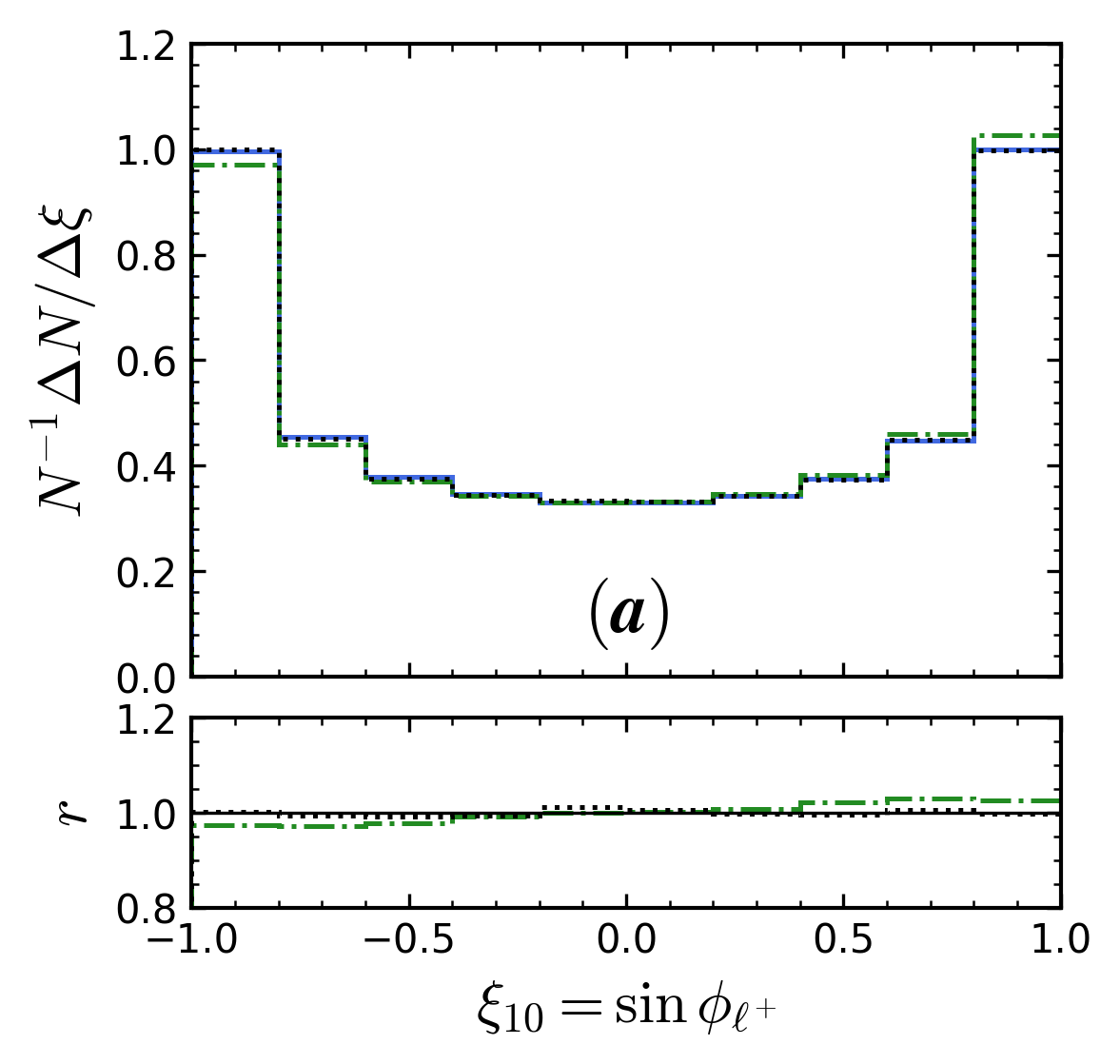}
\hfill
\includegraphics[width=0.31\textwidth]{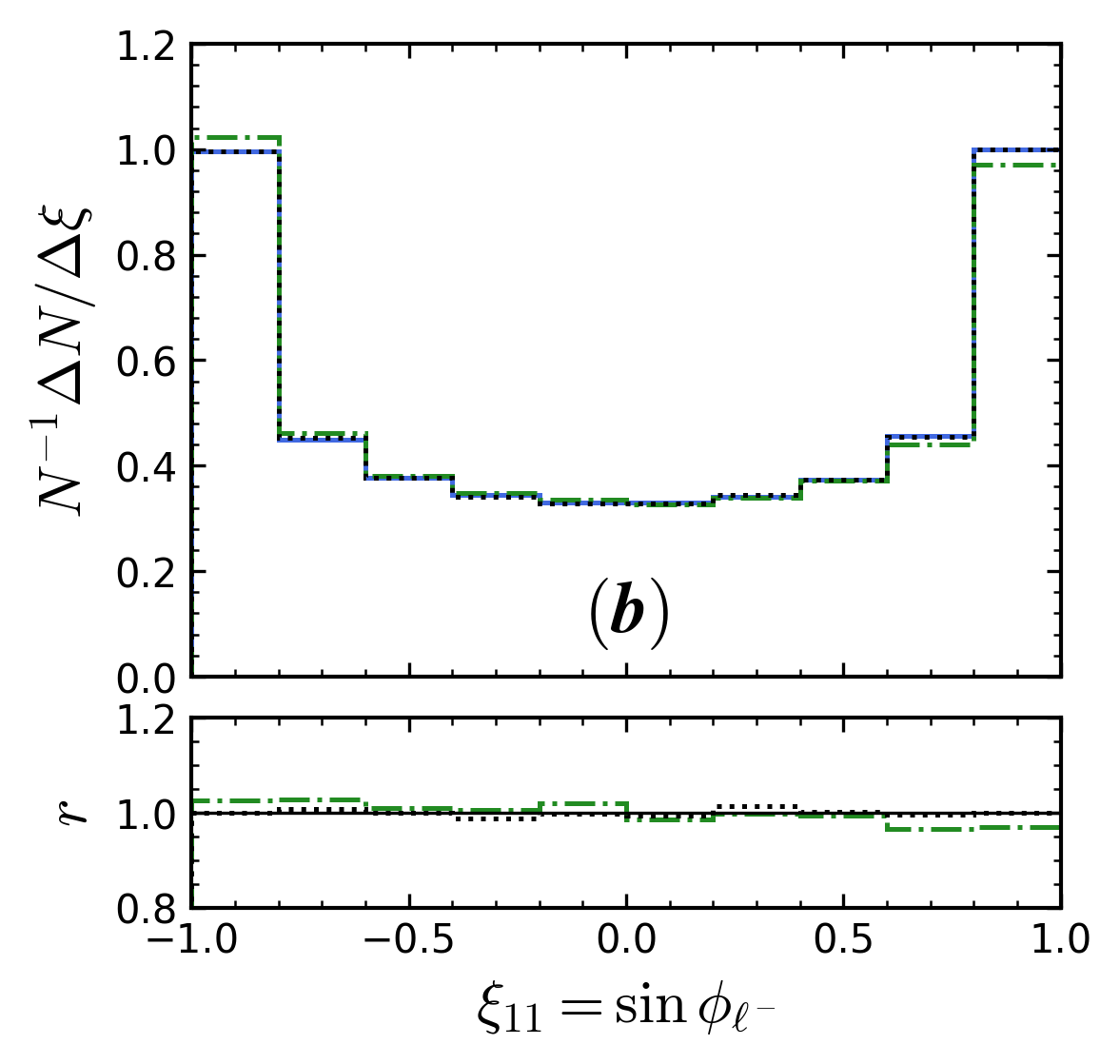}
\hfill
\includegraphics[width=0.31\textwidth]{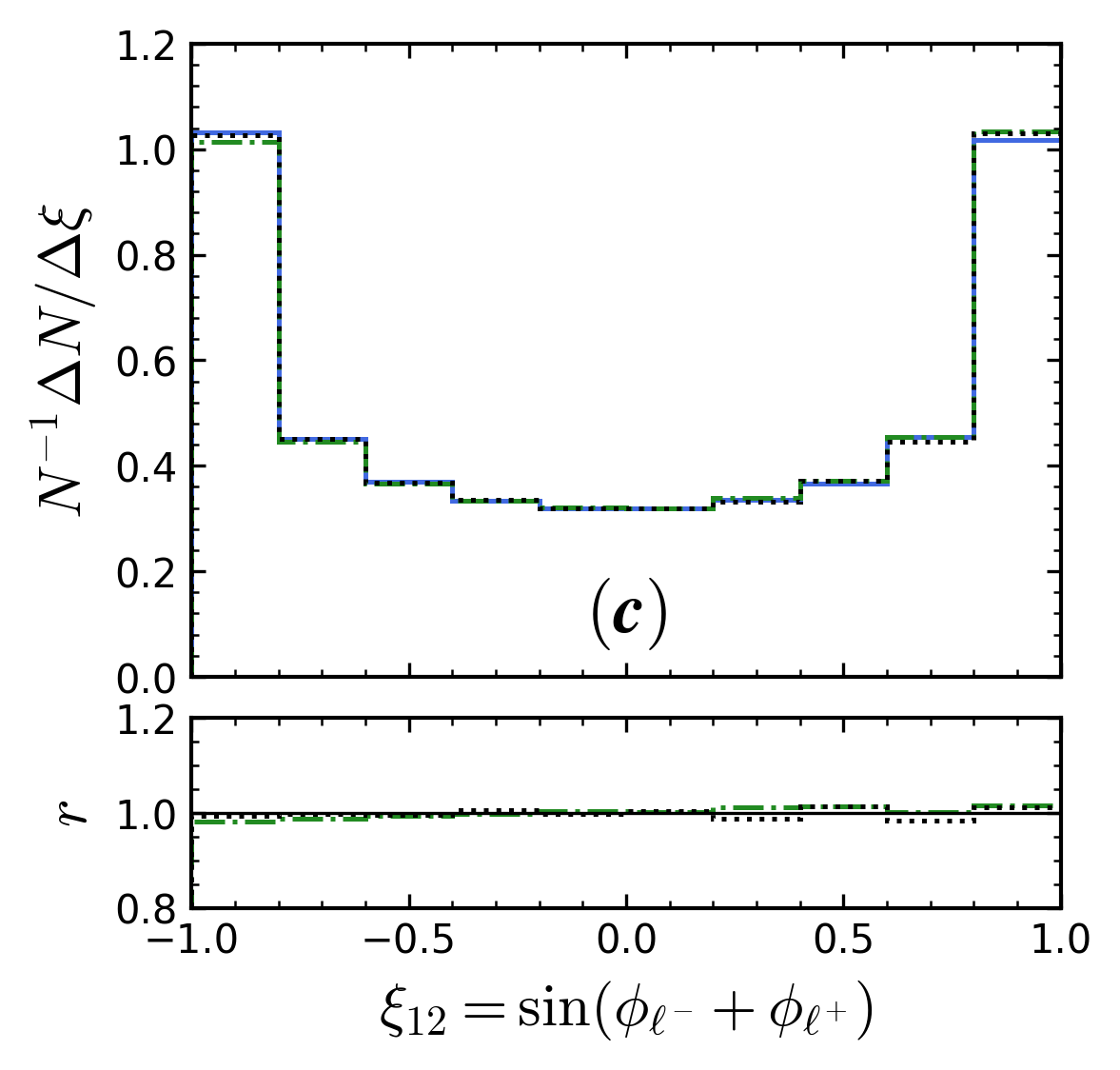}
\hfill
\includegraphics[width=0.31\textwidth]{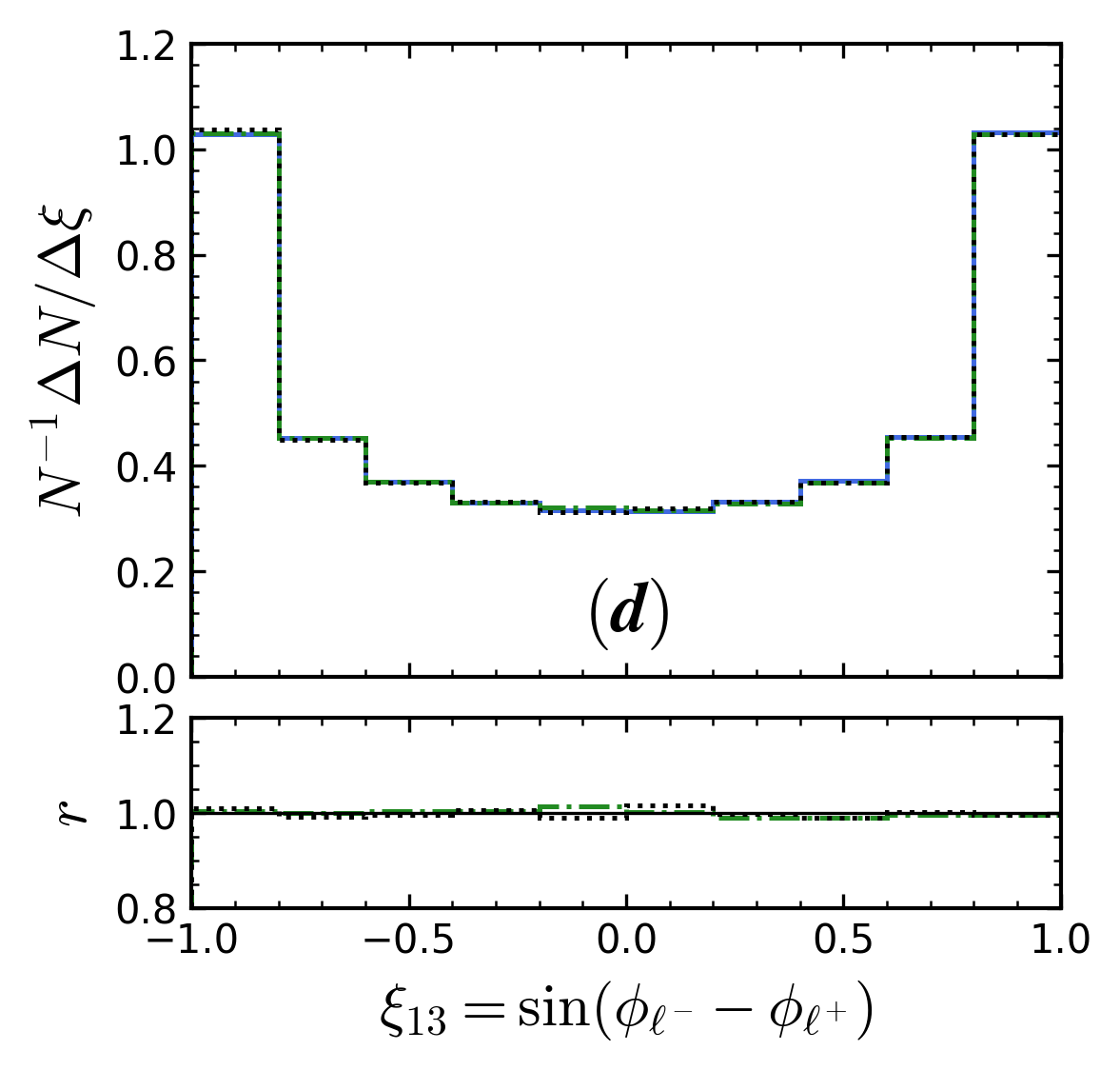}
\hfill
\includegraphics[width=0.31\textwidth]{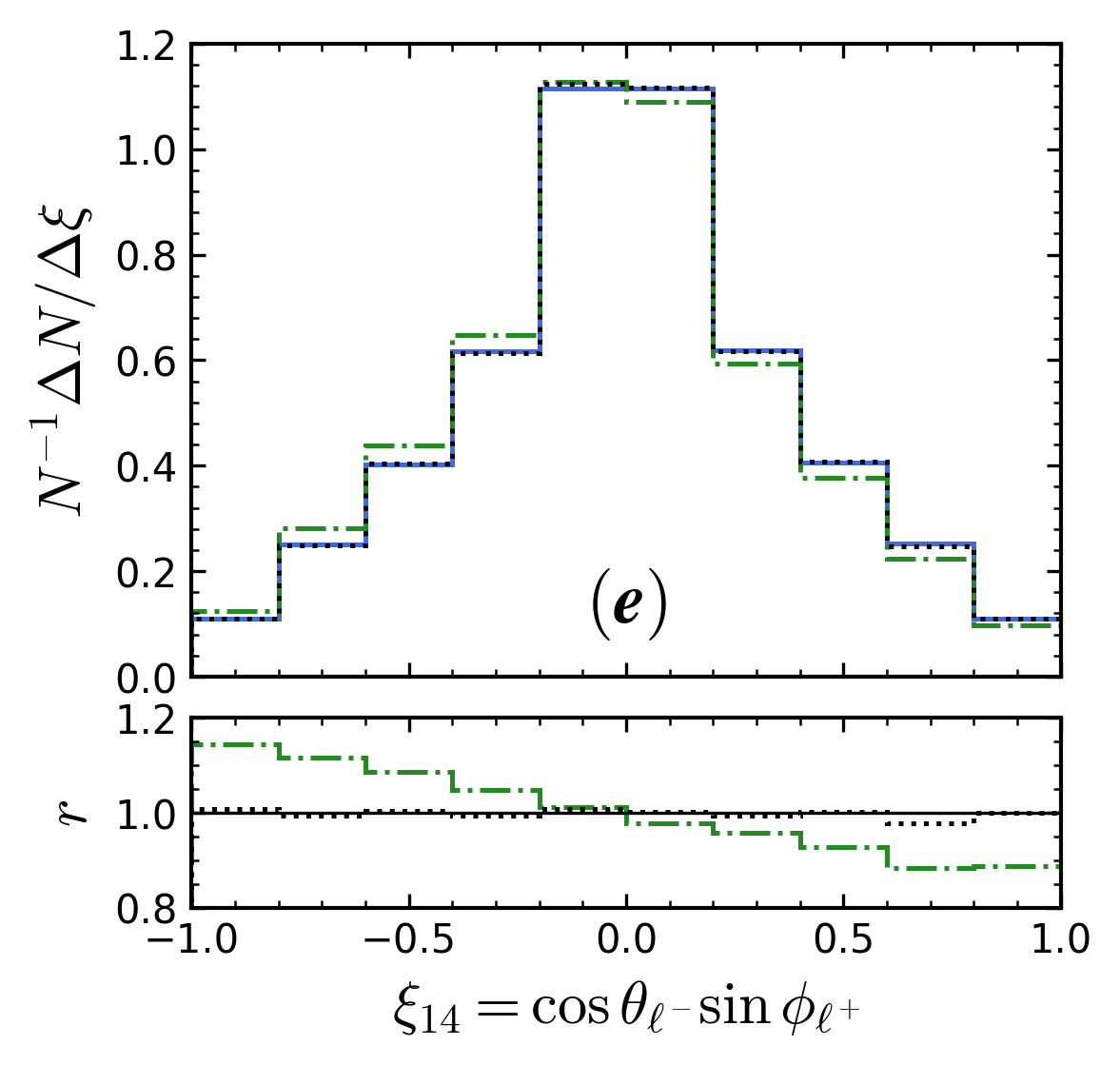}
\hfill
\includegraphics[width=0.31\textwidth]{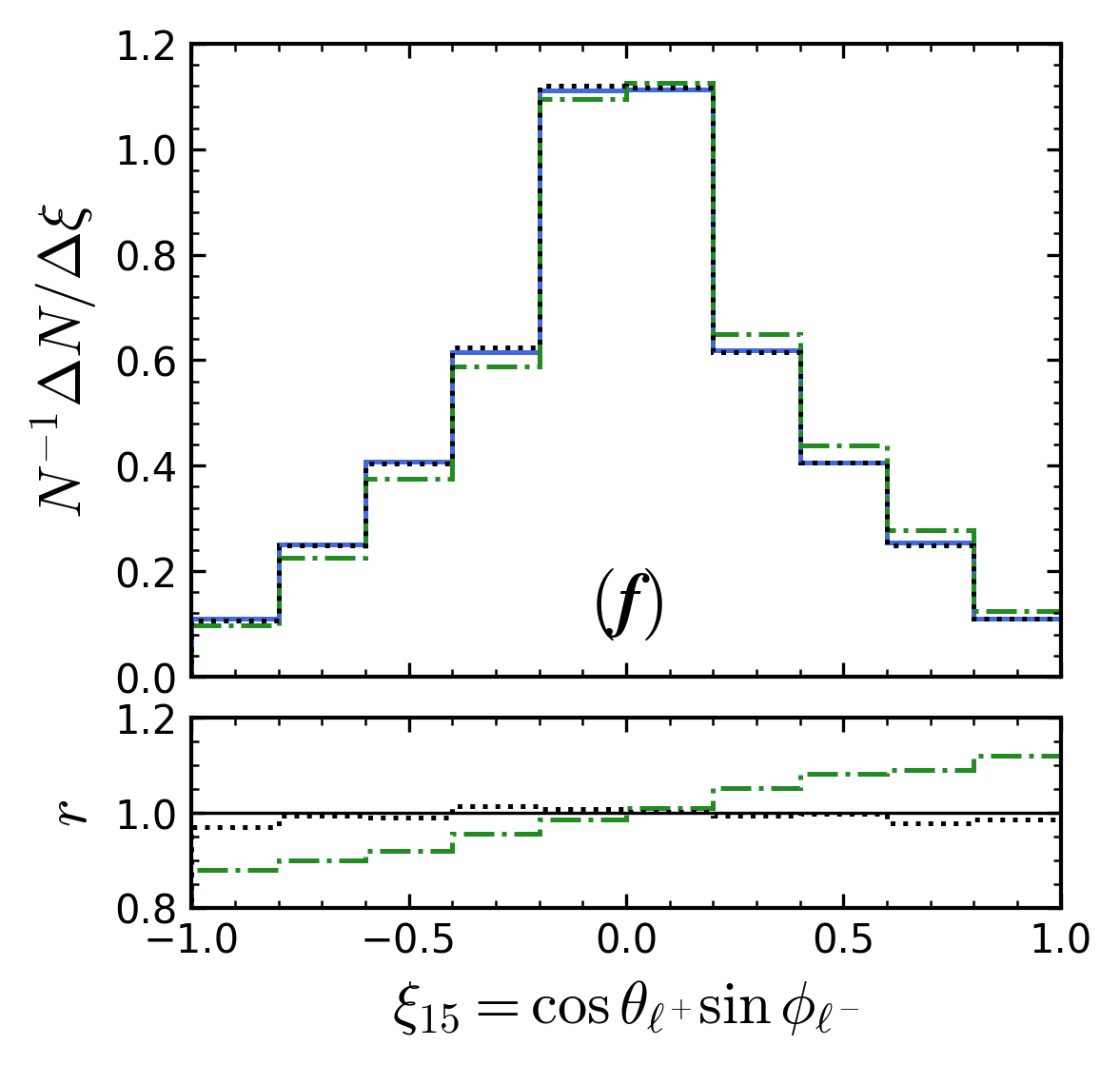}
\caption{\it
Normalized distributions of CP-odd lepton angular observables in the dileptonic process
$pp(gg)\to t(\ell^+\nu_\ell b)\bar t(\ell^-\bar\nu_\ell\bar b)$ at
$\sqrt{s}=13~{\rm TeV}$ in the near-threshold region
$330~{\rm GeV}<m_{t\bar t}<360~{\rm GeV}$.
The upper subpanels show the normalized distributions,
$N^{-1}\Delta N/\Delta\xi$, as functions of the angular variable $\xi$,
while the lower subpanels display the corresponding ratios $r$ with respect to
the SM $\eta_t$ prediction.
The CP-odd observables are
(a) $\xi_{10}=\sin\phi_{\ell^+}$,
(b) $\xi_{11}=\sin\phi_{\ell^-}$,
(c) $\xi_{12}=\sin(\phi_{\ell^-}+\phi_{\ell^+})$,
(d) $\xi_{13}=\sin(\phi_{\ell^-}-\phi_{\ell^+})$,
(e) $\xi_{14}=\cos\theta_{\ell^-}\sin\phi_{\ell^+}$, and
(f) $\xi_{15}=\cos\theta_{\ell^+}\sin\phi_{\ell^-}$.
The blue solid curves correspond to the SM color-singlet $\eta_t$ contribution,
the green dash-dotted curves show the corresponding prediction including the
top quark CEDM, and the black dotted curves denote the color-octet
${}^1S_0^{[8]}$ contribution.
}
\label{fig:csy}
\end{figure}

Figs.~\ref{fig:asy} and~\ref{fig:csy} show the normalized distributions of the lepton angular observables used to construct CP-even and CP-odd asymmetries, respectively, in the leptonic process \(pp(gg)\to t(\ell^+\nu_\ell b)\bar t(\ell^-\bar\nu_\ell\bar b)\) at \(\sqrt{s}=13~{\rm TeV}\). The distributions are evaluated in the near-threshold region with $330~{\rm GeV}<m_{t\bar t}<360~{\rm GeV}$ and normalized as \(N^{-1}\Delta N/\Delta\xi\), where \(\xi\) denotes the corresponding angular variable. The ratio \(r\) shown in the lower subpanels highlights deviations from the SM color-singlet \(\eta_t\) prediction and allows differences in the angular shapes to be compared independently of the overall normalization. The benchmark CEDM coupling is also taken as $d_t^g=3\times10^{-4}~{\rm GeV}^{-1}$.

For the CP-even observables in Fig.~\ref{fig:asy}, the normalized distributions are generally insensitive to the CEDM. The SM and SM+CEDM predictions for the \(\eta_t\) channel nearly overlap over the full angular range. The color-octet \({}^1S_0^{[8]}\) contribution also exhibits broadly similar angular dependence, although somewhat larger deviations are present for some observables. Since each distribution is normalized to its total event yield, this comparison isolates differences in the angular shapes rather than in the absolute production rates. Thus, although the different contributions can lead to sizable differences in the total cross sections, these differences are only weakly reflected in the shapes of the CP-even angular distributions. Consequently, the normalized CP-even observables provide only limited sensitivity to the CEDM in the near-threshold region.

A different behavior emerges for the CP-odd observables shown in Fig.~\ref{fig:csy}. While the single-lepton azimuthal variables and the two-lepton azimuthal combinations exhibit only small CEDM-induced modifications, the mixed polar--azimuthal correlations, \(\cos\theta_{\ell^-}\sin\phi_{\ell^+}\) and \(\cos\theta_{\ell^+}\sin\phi_{\ell^-}\), show a clear dependence on the CEDM. In particular, the CEDM produces opposite distortions on the two sides of the distributions, which can be efficiently captured by sign-weighted asymmetries. This behavior reflects the sensitivity of these observables to CP-violating spin correlations arising from the interference between the SM and CEDM-induced amplitudes. Since such interference contributions are linear in \(d_t^g\), CP-odd lepton angular asymmetries can provide substantially greater sensitivity to a small CEDM than the inclusive rate or the CP-even normalized distributions.

\begin{table}[htbp]
    \centering
    \caption{\it
    Approximate observed and predicted $t\bar t$ event yields in the three
    $c_{\mathrm{hel}}$ bins, extracted from Ref.~\cite{CMS:2025kzt} for
    $pp$ collisions at $\sqrt{s}=13~{\rm TeV}$ with an integrated luminosity
    of $138~{\rm fb}^{-1}$. $c_{\mathrm{hel}}$ is defined as the scalar product of the unit vectors of the two lepton momenta in the rest frames of their parent top quarks. The quoted uncertainties on the data yields are
    estimated from the corresponding figure.
    }
    \label{tab:upper_panel_data}
    \renewcommand{\arraystretch}{1.5}
    \begin{tabular}{cccc}
        \hline
        $c_{\mathrm{hel}}$ bin
        & Bin center
        & Data events
        & \quad $t\bar t$ events \\
        \hline
        $[-1,\,-1/3]$
        & $-0.67$
        & $2.42\times10^{4}\pm200$
        & \quad $2.16\times10^{4}$ \\

        $[-1/3,\,+1/3]$
        & $0$
        & $3.46\times10^{4}\pm250$
        & \quad $3.03\times10^{4}$ \\

        $[+1/3,\,+1]$
        & $+0.67$
        & $3.54\times10^{4}\pm300$
        & \quad $3.00\times10^{4}$ \\
        \hline
    \end{tabular}
\end{table}
The total near-threshold cross sections in the invariant mass window
$330~{\rm GeV}<m_{t\bar t}<360~{\rm GeV}$ are about $25.16~{\rm pb}$ and $29.27~{\rm pb}$ at the LHC13 and LHC14, respectively.
To estimate the number of selected events expected in our analysis, we normalize our
cross section predictions to the event yield reported by CMS in
Ref.~\cite{CMS:2025kzt}. The analysis is based on $pp$ collision data at
$\sqrt{s}=13~{\rm TeV}$ corresponding to an integrated luminosity of
$138~{\rm fb}^{-1}$. Approximate event yields extracted from the relevant distribution
are summarized in Tab.~\ref{tab:upper_panel_data}.
Summing over the three bins, we have $9.42 \times 10^4$ events in total, and $8.19\times10^4$ events from the $t\bar{t}$ channel.
Using average leptonic decay branching ratio $R_{t\to\ell} \approx 0.11$,
and comparing our cross section prediction with the
$t\bar t$ yield extracted from Ref.~\cite{CMS:2025kzt},
we can easily obtain the effective detector efficiency near the threshold, $\epsilon_{\rm tot} \simeq 0.487$.
Assuming the same total effective efficiency, the expected
number of selected events at LHC14 is therefore approximately
$9.53\times10^{4}$ for $138~{\rm fb}^{-1}$. For an integrated luminosity of
$3~{\rm ab}^{-1}$, representative of the HL-LHC data set, this yield increases to
approximately $2.07\times 10^{6}$ events.

\begin{figure}[t]
\centering
\includegraphics[width=0.58\textwidth]{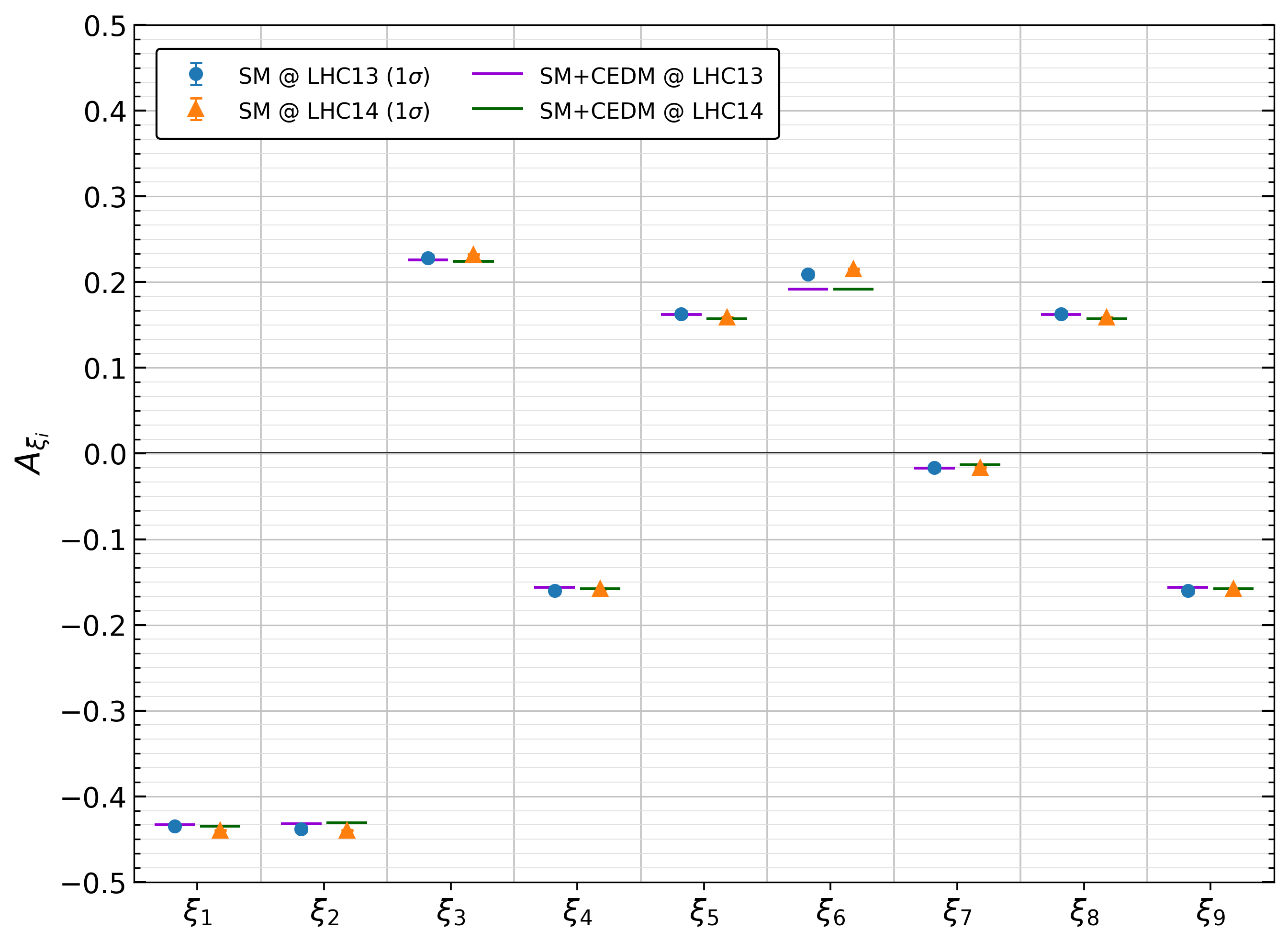}
\includegraphics[width=0.39\textwidth]{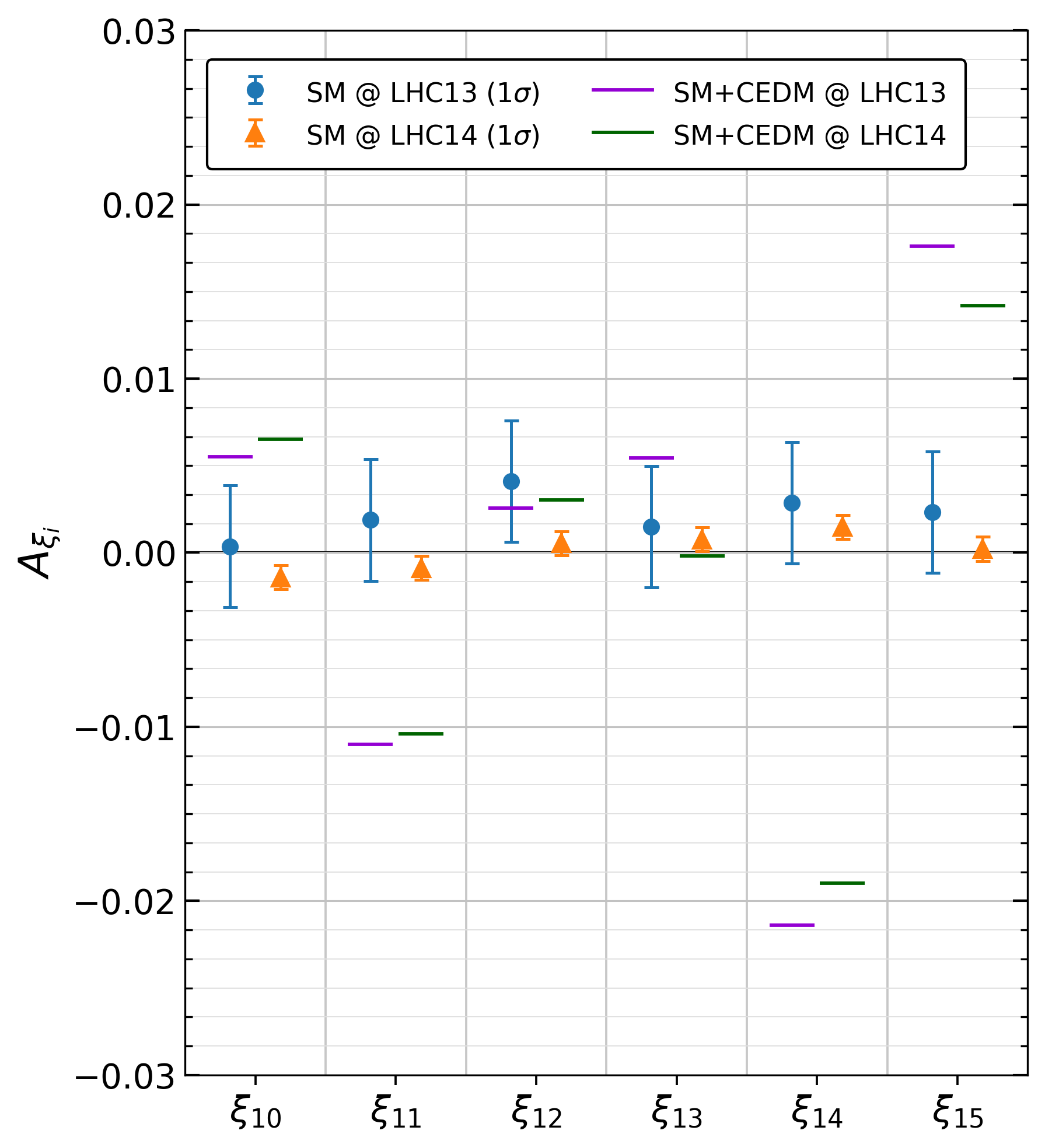}
\caption{\it
Predictions of the CP-even (left) and CP-odd (right) lepton angular asymmetry for observables $\xi_i$, in near-threshold $t\bar t$ production at the LHC.
The blue circles and orange triangles represent the SM predictions at LHC13 and LHC14, respectively, with error bars corresponding to the $1\sigma$ statistical uncertainties.
The integrated luminosities are $138~{\rm fb}^{-1}$ for LHC13 and $1~{\rm ab}^{-1}$ for LHC14.
The purple and green horizontal lines show the corresponding predictions including a top quark CEDM of $d_t^g=3\times10^{-4}~{\rm GeV}^{-1}$ at LHC13 and LHC14, respectively.
The definitions of the CP-even and CP-odd observables $\xi_i$ are given in the text.
}
\label{fig:XS:PP}
\end{figure}

Fig.~\ref{fig:XS:PP} summarizes the sensitivity of the CP-even and CP-odd lepton angular asymmetries to the top quark CEDM. For the CP-even observables, as shown in the left panel, the inclusion of \(d_t^g=3\times10^{-4}~{\rm GeV}^{-1}\) generally produces only modest shifts relative to the SM predictions, consistent with the weak CEDM dependence observed in the corresponding normalized angular distributions. A qualitatively different behavior is found for the CP-odd observables given in the right panel. In the SM, these asymmetries are expected to vanish or remain close to zero, whereas a nonzero CEDM can generate finite values through the interference between the SM and CEDM-induced amplitudes. The resulting separation between the SM and SM+CEDM predictions therefore provides a direct measure of the sensitivity to CP violation. The larger integrated luminosity assumed for LHC14 substantially reduces the statistical uncertainties, improving the sensitivity to these CEDM-induced deviations. These results demonstrate the complementarity between the two classes of observables: the CP-even asymmetries probe modifications of the spin-correlation structure, while the CP-odd asymmetries provide a more direct probe of the CP-violating interaction and retain linear sensitivity to \(d_t^g\).

\subsection{Prospective sensitivity}
\label{sec:sens}

In this section, we derive the expected exclusion limit on the top quark CEDM parameter $d_t^g$ at 13 TeV LHC, using the lepton angular asymmetries introduced above. Although a complete set of CP-even and CP-odd angular observables can be constructed from the leptonic final state, only a subset of them exhibit appreciable sensitivity to $d_t^g$ in the near-threshold region.
As shown in Fig.~\ref{fig:XS:PP}, five asymmetry observables exhibit particularly pronounced sensitivity to the CEDM and are therefore retained for the statistical analysis. We collect these observables into the vector
\begin{equation}
\mathbf{A}(d_t^g)
=
\left(
A_{\xi_6}(d_t^g),
A_{\xi_{10}}(d_t^g),
A_{\xi_{11}}(d_t^g),
A_{\xi_{14}}(d_t^g),
A_{\xi_{15}}(d_t^g)
\right)^T \;,
\label{eq:four_observables}
\end{equation}
where $\xi_6$ is CP-even, while $\xi_{10,11,14,15}$ are CP-odd. For the expected sensitivity analysis, we adopt an Asimov data set by taking the central values of the measured asymmetries to be their SM predictions, $\mathbf{A}_{\rm obs}\equiv\mathbf{A}_{\rm SM}=\mathbf{A}(0)$, which are given as,
\bea
A_{\xi_6}(0)&=& 0.212 \,,
\\[2mm]
A_{\xi_{10}}(0) &=& -1.45 \times 10^{-4} \,,
\\[2mm]
A_{\xi_{11}}(0) &=& -2.06 \times 10^{-4} \,,
\\[2mm]
A_{\xi_{14}}(0) &=& -5.00 \times 10^{-5} \,,
\\[2mm]
A_{\xi_{15}}(0) &=&  -1.40 \times 10^{-6}  \,.
\ena
One can see that the above asymmetries of CP-odd observables are not exactly zero.
This arises purely from statistical fluctuation, because the data set has a finite number of events. In our simulation, $10^7$ events are used to estimate the above expected values.
The expected number of selected events is
\begin{equation}
N_{\rm evt}
=
{\cal L}\,\sigma\,\epsilon_{\rm tot}\;,
\label{eq:expected_event_number_compact}
\end{equation}
where ${\cal L}$ is the integrated luminosity, $\sigma$ is the relevant cross section
including branching fractions, and $\epsilon_{\rm tot} \simeq 0.487$ denotes the total selection efficiency at LHC13.

Since all five asymmetries are evaluated from the same event sample, their statistical correlations must be retained.
The covariance matrix is defined by
\begin{equation}
V_{ij}
=
{\rm Cov}(A_{\xi_i},A_{\xi_j})\;,
\qquad
i,j=6,10,11,14,15\;.
\label{eq:covariance_four}
\end{equation}
In particular, the diagonal entries reduce to the usual binomial expression,
\begin{equation}
V_{ii}
=
\sigma_i^2
=
\frac{1-\left(A_{\xi_i}^{\rm SM}\right)^2}{N_{\rm evt}}\;.
\label{eq:asymmetry_error_compact}
\end{equation}
On the other hand, the off-diagonal elements, $V_{ij}$, are estimated with 5000 pseudo-experiments by repeatedly
resampling $N_{\rm evt}$ events from a large unweighted Monte Carlo parent sample $10^7$ events and
computing the five asymmetries in each pseudo-experiment,
\begin{equation}
V_{ij}
=
\frac{1}{N_{\rm toy}-1}
\sum_{k=1}^{N_{\rm toy}}
\left(A_{\xi_i}^{(k)}-\overline{A}_{\xi_i}\right)
\left(A_{\xi_j}^{(k)}-\overline{A}_{\xi_j}\right),
\qquad
\overline{A}_{\xi_i}=
\frac{1}{N_{\rm toy}}
\sum_{k=1}^{N_{\rm toy}} A_{\xi_i}^{(k)} \,,
\label{eq:pseudo_covariance_compact}
\end{equation}
where $A_{\xi_i}^{(k)}$ denotes the
$i$-th asymmetry in the $k$-th pseudo-experiment.
The same pseudo-experimental samples are used for all five observables, so that the
off-diagonal correlations are automatically included. For an expected limit around the
SM point, we use the covariance matrix evaluated at $d_t^g=0$.  This is appropriate as
long as the asymmetry uncertainties and event selection efficiency do not vary
significantly over the range of $d_t^g$ being tested.
The covariance matrix for LHC13 with an integrated luminosity of $138~{\rm fb}^{-1}$ is given as
\begin{equation}
\begin{pmatrix}
1.2\times 10^{-05} & -1.8\times 10^{-07} & 1.7\times 10^{-09} & 5.0\times 10^{-08} & -6.8\times 10^{-08} \\
-1.8\times 10^{-07} & 1.2\times 10^{-05} & 2.0\times 10^{-06} & -5.3\times 10^{-06} & -1.2\times 10^{-06} \\
1.7\times 10^{-09} & 2.0\times 10^{-06} & 1.2\times 10^{-05} & -1.3\times 10^{-06} & -5.3\times 10^{-06} \\
5.0\times 10^{-08} & -5.3\times 10^{-06} & -1.3\times 10^{-06} & 1.2\times 10^{-05} & 5.4\times 10^{-08} \\
-6.8\times 10^{-08} & -1.2\times 10^{-06} & -5.3\times 10^{-06} & 5.4\times 10^{-08} & 1.2\times 10^{-05}
\end{pmatrix}\;.
\end{equation}
One can see that there are relatively large anti-correlations between $\xi_{10}=\sin\phi_{\ell^+}$ and $\xi_{14}=\cos\theta_{\ell^-}\sin\phi_{\ell^+}$, and between $\xi_{11}=\sin\phi_{\ell^-}$ and $\xi_{15}=\cos\theta_{\ell^+}\sin\phi_{\ell^-}$. This is because each pair contains sine of the same azimuthal angle. As a result, their event-by-event fluctuations are strongly related, while the additional $\cos\theta_{\ell^\pm}$ weighting determines the negative sign.
These are the largest correlations.
A smaller but still visible correlation between $\xi_{10}=\sin\phi_{\ell^+}$ and $\xi_{11}=\sin\phi_{\ell^-}$ originates from the spin correlation between the top and anti-top quarks. It induces correlations between the angular distributions of their decay leptons.
In contrast, $\xi_6=\cos(\phi_{\ell^-}-\phi_{\ell^+})$ shows very small correlations with other observables. This is because they are different types of observables. Their interference therefore largely cancels when integrated over the approximately symmetric angular phase space. This makes them nearly orthogonal. Similarly, the very small correlation between $\xi_{14}$ and $\xi_{15}$ indicates substantial cancellation from the additional polar-angle weights, despite the underlying $t\bar t$ spin correlation.

The correlated chi-square function is then defined as
\begin{equation}
\chi^2(d_t^g)
=
\left[
\mathbf{A}_{\rm SM}
-
\mathbf{A}(d_t^g)
\right]^T
V_{\rm SM}^{-1}
\left[
\mathbf{A}_{\rm SM}
-
\mathbf{A}(d_t^g)
\right] \,.
\label{eq:chi2_four_observables}
\end{equation}
The exclusion interval is obtained from
\begin{equation}
\Delta\chi^2(d_t^g)
=
\chi^2(d_t^g)-\chi^2_{\min},
\qquad
\chi^2_{\min}
=
\min_{d_t^g}\chi^2(d_t^g)\;.
\label{eq:delta_chi2_four}
\end{equation}
Although five observables are included in the fit, only one model parameter, $d_t^g$, is
varied. Therefore the confidence interval on $d_t^g$ is determined using the
one-parameter $\Delta\chi^2$ criterion. The two-sided 95\% confidence interval is
defined by
\begin{equation}
\Delta\chi^2(d_t^g)
\leq
3.84\;.
\label{eq:dt_95CL_four}
\end{equation}
This procedure gives the combined expected exclusion limit from the five selected
asymmetry observables.

\begin{figure}[t]
\centering
\includegraphics[width=0.65\textwidth]{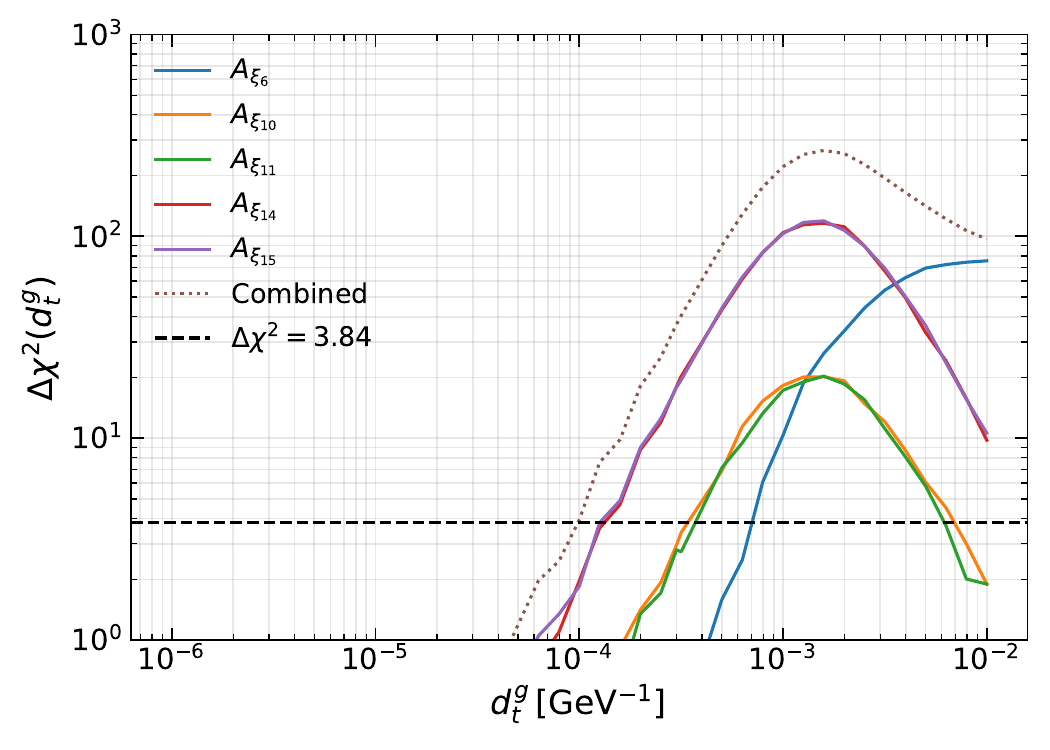}
\caption{
Expected $\Delta\chi^2$ curves as functions of the top quark CEDM parameter $d_t^g$, obtained from the five selected asymmetry observables. The curves labeled $A_{\xi_i}~(i=6,10,11,14,15)$ show the individual sensitivities of the five observables. The dotted curve represents their joint constraint, including correlations through the covariance matrix. The horizontal dashed line indicates $\Delta\chi^2=3.84$, corresponding to the 95\% CL.
}
\label{fig:asy:chi2}
\end{figure}

Fig.~\ref{fig:asy:chi2} shows the expected $\Delta\chi^2$ curves obtained from
the five asymmetry observables selected for their sensitivity to the dipole
parameter $d_t^g$. Each solid curve represents the sensitivity obtained
from an individual asymmetry observable from $A_{\xi_i}~(i=6,10,11,14,15)$. The dotted curve gives the joint
constraint obtained by consistently including the correlations among the
observables through the covariance matrix. At small values of $d_t^g$, the
$\Delta\chi^2$ values remain well below the 95\% CL threshold (horizontal dashed line). This
indicates that the predicted asymmetries cannot be statistically distinguished
from their SM expectations in this region. As $d_t^g$ increases, the
observables become increasingly sensitive to the CEDM parameter, although
the magnitude and functional dependence of this sensitivity differ among the
individual asymmetries because they probe different spin-correlation
structures.

The fluctuations visible at small $d_t^g$ and small $\Delta\chi^2$ originate from
the finite statistical precision of the simulated samples used to determine
the asymmetries and their covariance matrix. Since the underlying
$\Delta\chi^2$ values in this region are very small, modest absolute statistical
variations can appear relatively pronounced, particularly on the logarithmic
vertical scale used in Fig.~\ref{fig:asy:chi2}. These fluctuations should
therefore not be interpreted as physical structures in the dependence on $d_t^g$.
Their impact on the extracted bound is expected to be small provided that the
behavior of the $\Delta\chi^2$ curve in the vicinity of the confidence-level
threshold ($\Delta\chi^2=3.84$) is stable against variations of the Monte Carlo statistics.

A characteristic non-monotonic behavior is observed for the CP-odd
asymmetries $A_{\xi_i}~(i=10,11,14,15)$. Their $\Delta\chi^2$ values initially increase with
increasing $d_t^g$, reach a maximum, and subsequently decrease. This behavior can be
understood from the different powers of $d_t^g$ entering the interference and
squared CEDM contributions. For a normalized CP-odd asymmetry, the dependence
on $d_t^g$ can be written schematically as
\begin{equation}
    A_{\rm CP}(d_t^g)
    \sim
    \frac{c_1 d_t^g}{c_0 + c_2 d_t^{g2}}\;,
    \label{eq:acp-d-dependence}
\end{equation}
where the term proportional to $d_t^g$ in the numerator originates from the
interference between the SM and CEDM amplitudes, while $c_0$ and $c_2d_t^{g2}$
represent, respectively, the SM and squared-CEDM contributions to the
normalization. In the small $d_t^g$ region, where $c_2d_t^{g2} \ll c_0$, the asymmetry
is approximately linear in $d_t^g$, i.e.
$A_{\rm CP}(d_t^g)\sim (c_1/c_0)d_t^g$. As $d_t^g$ increases and the quadratic contribution
to the normalization becomes important, the growth of the asymmetry slows
down. Once $c_2d_t^{g2}$ dominates, the asymmetry scales approximately as
$A_{\rm CP}(d_t^g)\sim (c_1/c_2)/d_t^g$ and therefore decreases with increasing $d_t^g$.
This turnover provides a natural explanation for the rise and subsequent fall
of the $\Delta\chi^2$ curves associated with $A_{\xi_i}~(i=10,11,14,15)$. In contrast,
$A_{\xi_6}$ is CP-even, for which the leading CEDM-induced deviation is quadratic
in $d_t^g$. Its sensitivity therefore exhibits a qualitatively different behavior
and continues to increase over the range of $d_t^g$ shown in the figure.

The horizontal dashed line at $\Delta\chi^2=3.84$ corresponds to the
two-sided 95\% CL criterion for one parameter of interest.
To quote a single
conservative upper bound, we take the combined value of $d_t^g$ at which the
curve first crosses the threshold. This gives an expected upper bound of
approximately
\bee
    d_t^g \lesssim 9.6\times10^{-5} \, {\rm GeV}^{-1}
\ene
at the 95\% CL.
This is compatible to the current experimental upper bounds shown in Eq.~\eqref{eq:bul:CMS}.

\section{Conclusion}
\label{sec:Con}

The recent observation of an excess near the kinematic $t\bar{t}$ production threshold at the LHC implies the existence of a color-singlet and $S$-wave quasi-bound state of top quark. Due to the non-relativistic nature of top and anti-top quarks in the threshold region, the toponium gains the lowest momentum scale accessible in direct on-shell $t\bar t$ production. It is thus motivated to examine the anomalous dipole moments of top quark in toponium production.

In this work, we investigate the impact of CEDM in $t\bar{t}$-gluon vertex on toponium production at the LHC. We implement the enhanced near-threshold production of spin-0 color-singlet toponium state $\eta_t$ via gluon fusion. The subdominant productions of the spin-1 color-singlet state $\psi_t$ and color-octet states ${}^1S_0^{[8]}, {}^3S_1^{[8]}$ are also calculated for comparison. We combine the resonant toponium contribution
with the continuum $t\bar{t}$ production amplitude, including the interference effects induced by the top CEDM.

To probe the CP-violating interaction, we further construct a set of
asymmetries from fifteen angular distributions of the charged leptons in the leptonic decay channel of $t\bar{t}$. The observables defined in our method do not require the full reconstruction of the top and anti-top momenta. We then derive the expected exclusion limit on the top quark CEDM parameter $d_t^g$ using these lepton angular asymmetries. We find a 95\% CL upper bound $d_t^g \lesssim 9.6\times10^{-5} \, {\rm GeV}^{-1}$ at 13 TeV LHC.

\acknowledgments

We would like to thank Kaoru Hagiwara for helpful discussion.
T.~L. is supported by the National Natural Science Foundation of China (Grant No. 12375096). K. M. is supported by Shaanxi
Fundamental Science Research Project for Mathematicsand
Physics (Grant No. 25JSY031).

\bibliography{refs}

\end{document}